\documentclass[a4paper,fleqn, sort&compress]{cas-dc}
 \usepackage[numbers]{natbib}
\usepackage{graphicx}
\usepackage{lscape}
\usepackage{setspace}
\usepackage[toc,page]{appendix}
\usepackage{url}
\usepackage{hyperref}
\usepackage{booktabs}
\usepackage{multirow}
\usepackage{subfigure}
\usepackage{cuted}
\usepackage{mathtools}
\usepackage{amsmath}
\usepackage{amssymb}
\usepackage{lipsum}
\usepackage{bm}
\usepackage{color}
\usepackage{physics}
\usepackage{enumitem}
\usepackage{siunitx}
\newcommand{\iu}{\mathrm{i}\mkern1mu}

\newcommand{\appendixnumberline}[1]{Appendix\space}

\let\oldappendix\appendix
\makeatletter
\renewcommand{\appendix}{%
  \addtocontents{toc}{\let\protect\numberline\protect\appendixnumberline}%
  \renewcommand{\@seccntformat}[1]{Appendix~\csname the##1\endcsname.\space}%
  \oldappendix
}
\makeatother

\makeatletter
\newcommand*{\transpose}{\bgroup\@ifstar{\mathpalette\@transpose{\mkern-3.5mu}\egroup}{\mathpalette\@transpose\relax\egroup}}
\newcommand*{\@transpose}[2]{\setbox0=\hbox{\m@th$#1#2\intercal$}\raise\dp0\box0}
\makeatother

\begin{document}
\setlength\arraycolsep{2pt}

\let\WriteBookmarks\relax
\def\floatpagepagefraction{1}
\def\textpagefraction{.001}

\shorttitle{EMetaNode: Electromechanical Metamaterial Node for Broadband Vibration Attenuation and Self-powered Sensing}

\shortauthors{B. Zhao et~al.}

\title [mode = title]{EMetaNode: Electromechanical Friction-Induced    Metamaterial Node for Broadband Vibration Attenuation and Self-powered Sensing}

%

\author[1]{Bao Zhao}[orcid=0000-0002-9689-7742]
\credit{Conceptualization, Formal analysis, Investigation, Methodology, Resources, Writing – original draft}

\affiliation[1]{organization={Department of Civil and Environmental Engineering, Hong Kong Polytechnic University},
    city={Hung Hom, Kowloon},
    country={Hong Kong}}

\author[2]{Lorenzo Di Manici}
\credit{Validation, Investigation, Writing – review \& editing}

\author[2]{Raffaele Ardito}[orcid=0000-0002-4271-9190]
\credit{Resources, Writing – review \& editing}
\affiliation[2]{organization={Department of Civil and Environmental Engineering, Politecnico di Milano},
    city={Milan},
    postcode={20133},
    country={Italy}}

\author[3]{Eleni Chatzi}[orcid=0000-0002-6870-240X]
\credit{Supervision, Resources, Writing – review \& editing}
\affiliation[3]{organization={Department of Civil, Environmental, and Geomatic Engineering, ETH Z\"{u}rich},
    city={Z\"{u}rich},
    postcode={8093},
    country={Switzerland}}
    
\author[4]{Andrea Colombi}[orcid=0000-0003-2480-978X]
\credit{Supervision, Resources, Writing – review \& editing}
\cormark[1]
\ead{andrea.colombi@zhaw.ch}
\affiliation[4]{organization={Z\"urich University of Applied Sciences (ZHAW)},
    city={Winterthur},
    postcode={8401}, 
    country={Switzerland}}

\author[1]{Songye Zhu}[orcid=0000-0002-2617-3378]
\credit{Supervision, Resources, Writing – review \& editing}
\cormark[1]
\ead{songye.zhu@polyu.edu.hk}
\cortext[cor1]{Corresponding author}



\begin{abstract}
Recent advances in mechanical metamaterials and piezoelectric energy harvesting provide exciting opportunities for directing and converting mechanical energy in electromechanical systems for autonomous sensing and vibration control. However, practical realizations remain rare due to the lack of advanced modeling methods and persistent interdisciplinary barriers. By integrating mechanical metamaterials with power electronics-based interface circuits, this paper achieves a breakthrough by presenting an electromechanical friction-induced metamaterial node, which simultaneously enables self-powered sensing and broadband vibration attenuation. To support this, we introduce a reduced-order modeling framework combined with a numerical harmonic balance method tailored for nonlinear metamaterials. This approach efficiently captures local nonlinearities arising from electromechanical coupling through interface circuits, substantially improving computational efficiency. A key innovation of this work is that it uncovers the role of electromechanical friction, induced by synchronized switching interface circuits, which facilitates energy harvesting and enhanced nonlinear dynamic behavior—manifested through expanded bandgaps and higher-harmonic vibration attenuation. Experimentally, an electromechanical metamaterial node is realized for self-powered sensing of temperature and acceleration data, demonstrating strong potential for structural health monitoring and Internet of Things applications.
This study provides a practical pathway toward digitizing structures and systems by uniting smart interface circuitry with mechanical metamaterials to achieve autonomous, energy-aware sensing and control.

\end{abstract}



\begin{keywords}

electromechanical metamaterial \sep
electromechanical friction \sep
vibration attenuation \sep
energy harvesting \sep
self-powered sensing

\end{keywords}

\maketitle

\section{Introduction}

Ensuring autonomy, reliability, and efficiency in modern engineered systems increasingly depends on the ability to monitor and control mechanical responses in real time. This ability, in turn, demands effective guidance and conversion of mechanical energy—capabilities that can be enhanced by tailored materials and intelligent subsystems. Through electromechanical coupling (e.g., the piezoelectric effect), the shunted energy from mechanical systems can be regulated and harvested to power sensing components, enabling combined vibration attenuation and self-powered sensing for the Internet of Things (IoT) and autonomous structural health monitoring. 

Recent research efforts have revealed that locally resonant metamaterials are promising candidates to satisfy these demands. Unlike conventional single-degree-of-freedom tuned mass dampers \cite{Rahimi2020}, nonlinear vibration absorbers \cite{vakakis2001inducing}, and electromechanical shunting \cite{sodano2004review}, locally resonant metamaterials rely on the collective behavior of local resonators to tailor wave propagation in a frequency- and direction-dependent manner. This ability enables precise control over waveguiding and attenuation phenomena, offering a powerful platform for advanced vibration mitigation and energy manipulation.
Over the last decade, numerous studies \cite{craster2012acoustic,krushynska2023emerging} have focused on the mechanisms, designs, and fabrication of locally resonant metamaterials for attenuating low-frequency vibrations or amplifying incoming waves for energy harvesting or sensing purposes.  
Nevertheless, most existing research is limited to optimizing the mechanical part, e.g., optimizations of resonator designs and their arrangements \cite{krushynska2023emerging}, with little focus on electromechanical coupling dynamics. This gap becomes particularly critical when aiming to achieve dual functionality in vibration control and energy harvesting, especially under conditions of strong coupling. In such cases, the mechanical metamaterial and its external interface circuitry must be treated as an integrated system, where electrically induced effects—such as damping, stiffness, and inertia—can significantly influence wave propagation characteristics and system performance.

Recent studies have demonstrated that these integrated electromechanical metamaterials can lead to multiple functions beyond their original wave propagation tuning abilities, offering opportunities to realize programmable and multifunctional systems.  Through the piezoelectric coupling effect,  Airoldi et al. \cite{airoldi2011design}, Sugino et al. \cite{sugino2020digitally}, Silva et al. \cite{silva2020experimental}, Zheng et al. \cite{zheng2021adaptive}, and Jian et al. \cite{jian2023analytical} have demonstrated applications of active operational amplifiers to create equivalent impedance and transfer functions in the electrical domain for electrically programmable bandgap and vibration control. 
The applications of nonlinear shunting circuits \cite{silva2018experimentally,raze2019digital} and nonlinear synthetic impedance circuits \cite{alfahmi2024programmable} in electromechanical metamaterials enable further options for amplitude-dependent bandgap \cite{mosquera2021dynamics} and broadband vibration attenuation \cite{zhang2021metamaterial,chen2024bandgap,xia2024piezoelectric}. These advancements significantly enhance the wave manipulation capabilities of metamaterials, enabling more adaptive and efficient control over the propagation of mechanical energy across a wider range of operating conditions.
In addition, passive components such as the resistor–inductor–capacitor (RLC)  circuit branches are widely used in locally resonant metamaterials for energy harvesting \cite{alshaqaq2020graded,de2020graded} and vibration attenuation obtained via shunting \cite{zhou2015vibration,chen2016adaptive}. In these systems, electrically induced damping and stiffness can be explicitly quantified, providing valuable insight into how electromechanical dynamics affect wave propagation and bandgap formation. However, because these circuits are primarily analog, they face key limitations: They either rely on external power sources or devices for dynamic tuning, or they generate alternating current (AC) for energy harvesting. Such limitations restrict their practical applicability in scenarios where direct current (DC) power is needed to operate external devices, sensors, or IoT nodes.

To overcome these issues, power electronics-based piezoelectric interface circuits, such as synchronized electrical charge extraction (SECE) \cite{lefeuvre2005piezoelectric}, series synchronized switching on the inductor (S-SSHI) \cite{guyomar2011recent}, and series synchronized triple bias-flip (S-S3BF) \cite{zhao2020series},  have been proposed and applied in single degree of freedom systems for simultaneous energy harvesting and vibration control. With energy-efficient diodes and metal–oxide–semiconductor field-effect transistors (MOSFETs), self-powered regulation of piezoelectric voltage and DC power output can be achieved. As a result, they are strong candidates for autonomous electromechanical systems with the dual functions of energy harvesting and vibration control. However, the applications of these advanced interface circuits in electromechanical metamaterials remain limited, with only a few recent studies exploring their potential. Chen et al. \cite{chen2020elastic} and Zhao et al. \cite{zhao2022graded} utilized SECE circuits in locally resonant piezoelectric metamaterials to convert mechanical vibrations into DC voltage. Asanuma et al. investigated the damping effect with a variation of the S-SSHI circuit for the modal frequency attenuation of a metamaterial beam \cite{asanuma2022modal}. Bao et al. studied multimodal \cite{bao2020manipulating} and broadband attenuation \cite{bao2025piezoelectric} in metamaterial beams and plates. A typical modeling approach in these studies involves the use of impedance analysis \cite{liang2011impedance,zhao2023circuit} to obtain the linear equivalent impedance, which subsequently leads to electromechanically coupled dynamics. However, classical impedance analysis linearizes circuit behavior and overlooks the fundamentally nonlinear dynamics introduced by diodes and switching transistors \cite{banerjee1999nonlinear}. A key challenge, therefore, is to model and quantify these nonlinear effects and their impact on wave propagation, vibration attenuation, and energy harvesting.

The challenge of integrating power electronics with wave dynamics for vibration control and self-powered sensing lies in the nonlinear modeling of interface circuits with metamaterials and in realizing an autonomous electromechanical metamaterial system that converts, harvests, and regulates mechanical energy for sensing functions. In this study, we devise a reduced-order harmonic-balance (ROM–HB) method grounded in nonlinear frequency domain dynamics modeling \cite{krack2019harmonic,seydel2009practical}. This approach enables the treatment of coupled electromechanical dynamics with switching-based interface circuits by representing their effect as an equivalent electromechanical friction—a conceptual advance presented here for the first time. Combined with the inhomogeneous wave correlation (IWC) method \cite{van2018measuring}, this framework reveals the emergence of nonlinear band structures within an electromechanical metamaterial and their dependence on the input acceleration across various switching-based interface circuits.  Notably, the analysis highlights broadband attenuation and higher-harmonic-induced vibration attenuation, stemming from the nonlinear electromechanical friction effects. From an experimental perspective, electromechanically coupled wave propagation and attenuation, DC power performance, and experimental waveforms are reported and verified against theoretical studies. In addition, we propose a compact IoT board comprising an SECE interface circuit, an energy management circuit, and a low-power Bluetooth beacon chip with external sensors, enabling fully self-powered temperature and acceleration sensing. Together with the metamaterial, this electromechanical metamaterial node (EMetaNode) can achieve simultaneous broadband self-powered sensing and vibration control, providing new theoretical perspectives and practical design diagrams for fully autonomous structures and systems.

\section{Theoretical analyses}
\label{sec:theory}

While recent studies have explored various mechanical nonlinearities in metamaterials, including geometric nonlinearity \cite{bae2022nonlinear}, nonlinear damping \cite{zhao2024nonlinear,xiao2025broadband}, cubic stiffness \cite{gong2023band,li2025significantly}, contact, and vibro-impact \cite{fang2022nonlinear,gong2024multi}, the modeling of nonlinear electromechanical metamaterials remains a significant challenge due to the complex nonlinear dynamics, stability issues, and electromechanical coupling effects. Most existing research employs time domain methods \cite{manktelow2011multiple,narisetti2010}, which are broadly applicable and capable of capturing both periodic and aperiodic wave propagation. However, when the primary focus is on the frequency domain steady-state response, time-domain methods often become cumbersome, requiring significant computational effort for each desired frequency of interest. In addressing this issue, frequency domain-based methods such as the harmonic balance method \cite{krack2019harmonic,soleimani2025nonlinear} offer distinct advantages due to their fast convergence rate and predefined harmonic solutions. 
 In this work, we base our model on a finite element formulation of the underlying linear system, with nonlinearities localized in the electromechanical interface circuits coupled to the resonators. We solve the resulting system using the proposed ROM-HB method. Since the nonlinear degrees of freedom—originating from circuit–resonator interactions—are relatively limited compared to the full system, this framework enables efficient computation by confining the nonlinear dynamics to a reduced-order subspace. This approach significantly lowers computational costs while maintaining accuracy in capturing the essential nonlinear behavior of the electromechanical metamaterial.

\begin{figure*}[!t]
    \centering
    \includegraphics[width=2\columnwidth,page=1]{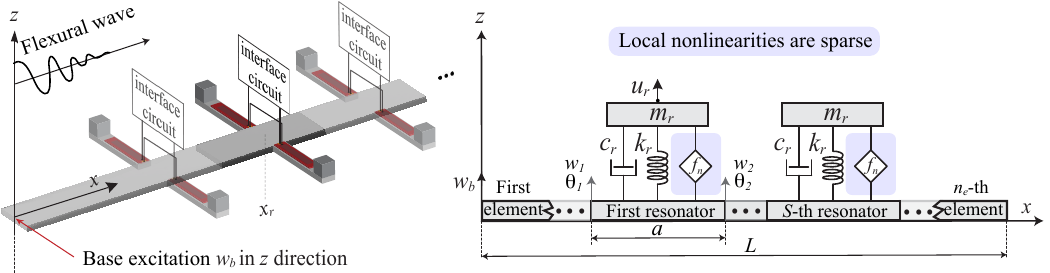}
    \caption{Schematic of the nonlinear metamaterial system and its lumped parameter model. The local nonlinearities of local resonators are represented as a general nonlinear reaction force $f_n$.}
    \label{fig:model}
\end{figure*}

\subsection{Reduced-order modeling of a nonlinear metabeam}
\label{sec:rom}

As shown in Fig.~\ref{fig:model}, the proposed nonlinear metamaterial consists of a linear host beam and multiple periodic pairs of piezoelectric parasitic beams with interface circuits. Each parasitic beam contains a cantilever beam attached to the host beam, a stainless steel tip mass, and two series-connected piezoelectric patches. 
The governing equation of the metamaterial beam can be represented via the Euler--Bernoulli beam equation, augmented with distributed reaction forces at the root position of each pair of parasitic beams
\begin{equation}
D_0 \frac{\partial^4 w}{\partial x^4}+\rho_0 \frac{\partial^2 w}{\partial t^2}=-\rho_0 \ddot{w}_b+2\delta(x-x_r)\sum_{r=1}^S  m_r\left(\ddot{u}_r+\ddot{w}+\ddot{w}_b\right), 
\label{eq:nlmeta1}
\end{equation}
where $D_0$ and $\rho_0$ are the rigidity modulus and linear density of the host beam, respectively. $m_r$ is the mass of the $r$-th nonlinear local resonator at position $x_r$ along the host beam. $w$ and $u_r$ represent the relative displacement of the host beam with respect to its clamped side and the relative displacement of local resonators with respect to the host beam, respectively. $w_b$ is the displacement of the clamped side due to base excitation.  The two piezoelectric patches in each pair of identical parasitic beams are connected in series; thus, each local resonator shares half of the nonlinear reaction force. The first bending mode of the parasitic beams \cite{hu2019modelling} can be used to derive the governing equation of a nonlinear local resonator, which reads
\begin{equation}
m_r \ddot{u}_r + c_r \dot{u}_r +k u_r + 0.5*f_n(u_r,\dot{u}_r)=-m_r \left(\ddot{w}+\ddot{w}_b\right),
\label{eq:nlmeta1_1}
\end{equation}
where the reaction force due to electromechanical coupling is a nonlinear function $f_n$ of the relative displacement and velocity $u_r$ and $\dot{u}_r$. $c_r$ and $k_r$ are the damping and stiffness of the $r$-th nonlinear local resonator. As a result of the nonlinear reaction force of the local resonators, the overall metamaterial system, as described by Eqs.~\eqref{eq:nlmeta1} and \eqref{eq:nlmeta1_1}, will also exhibit nonlinearities.

The underlying linear finite element model of the host beam is assembled using one-dimensional two-node beam elements, allowing for translational and rotational nodal degrees of freedom. With the use of cubic shape functions derived via Hamilton’s principle, the elemental mass, stiffness, and nodal force matrices are documented in Appendix \ref{appendixA}. By assembling $n_e$ elements, we can model the host beam of length $L$ with the dimension of $(2n_e+2)\times (2n_e+2)$.
As shown in Fig.~\ref{fig:model}, assuming the unit cell length $a$ of the metamaterial beam is discretized with elements of length $l_e$, we can introduce the sum of nonlinear reaction forces of a pair of local resonators at the translational node of an element. 
Adding $S$ nodes representing the number of pairs of parasitic beams obtains the total dimension of the global mass and stiffness matrices as $(2n_e+2+S)\times (2n_e+2+S)$. Each unit cell can be meshed with multiple beam elements. However, the mass and stiffness matrices of a single element unit cell are documented in Appendix \ref{appendixA} to demonstrate the assembly of the metamaterial system. 

As shown in Fig.~\ref{fig:model}, a single-element unit cell is formed with a nonlinear local resonator attached to a host beam segment with the unit cell length $a=l_e$.
The unknown displacement vector of the single-element unit cell is given as $\mathbf{q}_{\mathrm{er}}=\left[w_1, \theta_1, u_r, w_2, \theta_2 \right]^\transpose$, where the subscripts $1$ and $2$ represent the first and second nodes of the element, respectively. The columns and rows corresponding to the nonlinear local resonator are moved to the bottom-right corner of the global assembly of the metamaterial beam. The linear damping ratio $\zeta=0.002$ is applied to the local resonators and the host beam assembly, with Rayleigh damping determined for low- and high-frequency modes. The nonlinear finite element model of the electromechanical metamaterial beam reads
\begin{equation}
\mathbf{M}\mathbf{\ddot{q}}+\mathbf{C}\mathbf{\dot{q}}+\mathbf{K}\mathbf{q}+\mathbf{F}_{\mathrm{n}}=\mathbf{F}_{\mathrm{ex}}, 
\label{eq:fem}
\end{equation}
in which
\begin{equation}
\begin{array}{lll}
\mathbf{M}=\left[\begin{array}{cc}
\mathbf{M}_{\mathrm{LL}} & \mathbf{M}_{\mathrm{LN}} \\
\mathbf{M}_{\mathrm{NL}} & \mathbf{M}_{\mathrm{NN}}
\end{array}\right],  & 
 \mathbf{C}=\left[\begin{array}{cc}
\mathbf{C}_{\mathrm{LL}} & \mathbf{C}_{\mathrm{LN}} \\
\mathbf{C}_{\mathrm{NL}} & \mathbf{C}_{\mathrm{NN}}
\end{array}\right],  & 
 \\
\rule{0pt}{5ex} 
\mathbf{K}=\left[\begin{array}{cc}
\mathbf{K}_{\mathrm{LL}} & \mathbf{K}_{\mathrm{LN}} \\
\mathbf{K}_{\mathrm{NL}} & \mathbf{K}_{\mathrm{NN}}
\end{array}\right], &
\mathbf{F}_{\mathrm{n}}=\left[\begin{array}{c}
\mathbf{0}  \\
\mathbf{F}_{\mathrm{n}}(\mathbf{q}_{\mathrm{N}},\mathbf{\dot{q}}_{\mathrm{N}})
\end{array}\right], &
\mathbf{q}=\left[\begin{array}{c}
\mathbf{q}_{\mathrm{L}}  \\
\mathbf{q}_{\mathrm{N}}
\end{array}\right], 
\end{array}
\end{equation}
where the matrices $\mathbf{M}_{\mathrm{LL}}$, $\mathbf{C}_{\mathrm{LL}}$, $\mathbf{K}_{\mathrm{LL}}$, and $\mathbf{q}_{\mathrm{L}}$ correspond to the mass, damping, and stiffness matrices, as well as the unknown displacement vector of the host linear beam. $\mathbf{M}_{\mathrm{LN}}=\mathbf{M}_{\mathrm{NL}}^T$ contains the masses of the nonlinear local resonators that introduce the reaction forces on the host beam and the local resonators, as indicated by the right-hand sides of Eqs.~\eqref{eq:nlmeta1} and \eqref{eq:nlmeta1_1}. Thus, the off-diagonal submatrices in $\mathbf{C}$ and $\mathbf{K}$ are zero matrices. $\mathbf{M}_{\mathrm{NN}}$, $\mathbf{C}_{\mathrm{NN}}$, $\mathbf{K}_{\mathrm{NN}}$, $\mathbf{q}_{\mathrm{N}}$, and $\mathbf{F}_{\mathrm{n}}(\mathbf{q}_{\mathrm{N}},\mathbf{\dot{q}}_{\mathrm{N}})$ represent the mass matrix, damping matrix, stiffness matrix, unknown displacement vector, and the nonlinear reaction force vector of the nonlinear local resonators, respectively. The first row of Eq.~\eqref{eq:fem} is the governing equation for the linear host beam with a dimension of $(2n_e+2)$. The second row of Eq.~\eqref{eq:fem} represents the governing equations for nonlinear resonators with dimension $S$.

\begin{figure*}[!t]
    \centering
    \includegraphics[width=2\columnwidth,page=2]{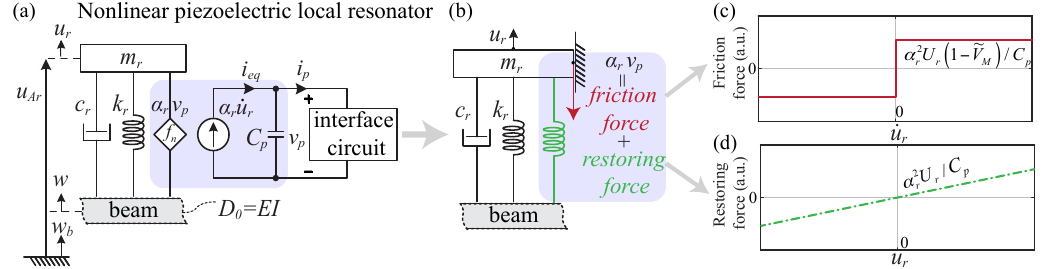}
     \caption{Schematic of the nonlinear local resonator and its reaction forces from the piezoelectric interface circuit. (a) Piezoelectric local resonator with an interface circuit; (b) Equivalent nonlinear local resonator with the nonlinear friction and restoring forces from the switching-based interface circuits; (c) Relationship between the friction force and the velocity $\dot{u}_r$; (d) Relationship between the restoring force and the displacement $u_r$. }
    \label{fig:resonator}
\end{figure*}

By treating the unknown displacements as a Fourier series of multiple harmonics and substituting the ansatz of $\mathbf{q}(t)$: $\mathbf{q}(\hat{\mathbf{q}}_H,t)$ into Eq.~\eqref{eq:fem} up to the truncation order $H$, we can form the frequency domain representation of the nonlinear metamaterial beam. The representation for the $k$-th harmonic order is given as
\begin{equation}
\mathbf{S}^k(\omega)\hat{\mathbf{q}}^k + 
\hat{\mathbf{F}}^k_{\mathrm{n}}\left(\hat{\mathbf{q}}^k, \omega\right) =
\hat{\mathbf{F}}_{\mathrm{ex}}(\omega),
\label{eq:hbres}
\end{equation}
in which
\begin{equation}
\mathbf{S}^k(\omega)=-k^2 \omega^2 \mathbf{M} + \iu k \omega \mathbf{C}+  \mathbf{K},
\end{equation}
where $k=-H,\ldots,H$ is the index of the harmonic order with dimension $2H+1$, and $\hat{\square}$ represents the complex Fourier coefficient. 

Since the nonlinear reaction forces arise from the local resonators rather than the host beam, they can be regarded as local nonlinearities, and Eq.~\eqref{eq:hbres} can be rewritten with the residual function $\mathbf{r}^k(\hat{\mathbf{q}}^k,\omega)$ as
\begin{equation}
\mathbf{r}^k(\hat{\mathbf{q}}^k,\omega)=   \mathbf{H}^k(\omega)                
\left(\left[\begin{array}{c}
\mathbf{0}\\
\hat{\mathbf{f}}_{\mathrm{n}}^k\left(\hat{\mathbf{q}}_{\mathrm{N}}, \omega\right) 
\end{array}\right]-\hat{\mathbf{F}}_{\mathrm{ex}}^k(\omega)\right)
\rule{0pt}{5ex}  
\left[\begin{array}{c}
\hat{\mathbf{q}}_{\mathrm{L}}^k \\
\hat{\mathbf{q}}_{\mathrm{N}}^k
\end{array}\right] =\mathbf{0}
\label{eq:condense1}
\end{equation}
in which
\begin{equation}
    \mathbf{H}^k(\omega)=[\mathbf{S}^k(\omega)]^{-1}=\left[\begin{array}{cc}
\mathbf{H}_{\mathrm{LL}}^k & \mathbf{H}_{\mathrm{LN}}^k \\
\mathbf{H}_{\mathrm{NL}}^k & \mathbf{H}_{\mathrm{NN}}^k
\end{array}\right] (\omega),
\end{equation}
where $\mathbf{H}^k(\omega)$ represents the dynamic compliance
matrix for the $k$-th harmonic. The first row of Eq.~\eqref{eq:condense1} corresponds to linear equations for the host beam, while the second row corresponds to the nonlinear equations for the nonlinear local resonators. We can therefore lump the nonlinearity at the coordinates of the nonlinear local resonators and retrieve the linear solutions with the solved nonlinear displacements. The condensed problem concerning the nonlinear local resonators can be stated as
\begin{equation}
    \mathbf{r}^k_{\mathrm{N}}(\hat{\mathbf{q}}_{\mathrm{N}}^k,\omega)=\hat{\mathbf{q}}_{\mathrm{N}}^k+\mathbf{H}_{\mathrm{NN}}^k \hat{\mathbf{f}}_{\mathrm{n}}^k\left(\hat{\mathbf{q}}_{\mathrm{N}}, \omega\right) - \mathbf{H}_{\mathrm{ex}}^k
\hat{\mathbf{F}}_{\mathrm{ex}}^k(\omega)=\mathbf{0}.
\label{eq:condense2}
\end{equation}
in which
\begin{equation}
\mathbf{H}_{\mathrm{ex}}^k=\left[\begin{array}{cc}
\mathbf{H}_{\mathrm{NL}}^k & \mathbf{H}_{\mathrm{NN}}^k
\end{array}\right].
\end{equation}

Compared with Eq.~\eqref{eq:condense1}, the condensation in Eq.~\eqref{eq:condense2} reduces the dimensionality of the problem from $(2n_e+2+S)(2H+1)$ to $S(2H+1)$, significantly boosting computational efficiency by separately solving the equations involving the nonlinear local resonators.
To solve the reduced-order problem described in Eq.~\eqref{eq:condense2}, an iterative Newton method is applied to gradually reduce the residual and achieve a local minimum at a given frequency $\omega$. The frequency $\omega$ is also appended as an element of the unknown to carry out the arc-length-based numerical continuation \cite{seydel2009practical,krack2019harmonic} for the frequency response solution within the frequency range $\left[\omega_s,\omega_e\right]$ (see Appendix \ref{appendixB} for details). The stability of the solutions can also be determined by perturbing the periodic solutions for Eq.~\eqref{eq:hbres} with Hill's method along the continuation trajectory \cite{von2001harmonic}. 
Solution solving and numerical computation are facilitated by presenting the Jacobian of Eq.~\eqref{eq:condense2} concerning $\hat{\mathbf{q}}_{\mathrm{N}}$ and $\omega$ as
\begin{equation}
\begin{aligned}
        \partial \mathbf{r}_{\mathrm{N}} &=\partial \hat{\mathbf{q}}_{\mathrm{N}}+ \partial \hat{\mathbf{f}}_{\mathrm{n}}\left(\hat{\mathbf{q}}_{\mathrm{N}}, \omega\right)\mathbf{H}_{\mathrm{NN}}(\omega)-\partial \hat{\mathbf{F}}_{\mathrm{ex}}(\omega) \mathbf{H}_{\mathrm{ex}}(\omega) \\ 
        & + \left(\partial \mathbf{H}_{\mathrm{NN}}(\omega) \hat{\mathbf{f}}_{\mathrm{n}}\left(\hat{\mathbf{q}}_{\mathrm{N}}, \omega\right)-\partial \mathbf{H}_{\mathrm{ex}}(\omega) \hat{\mathbf{F}}_{\mathrm{ex}}(\omega)\right)\mathrm{d}\omega.
\end{aligned}
    \label{eq:dR}
\end{equation}

The computational hurdle related to the Jacobian lies in the calculation of $\partial \mathbf{H}(\omega) $ and $\partial \hat{\mathbf{f}}_{n}\left(\hat{\mathbf{q}}_{\mathrm{N}}, \omega\right)$. Given that the matrices $\mathbf{M}$, $\mathbf{C}$, and $\mathbf{K}$ in Eq.~\eqref{eq:fem} are symmetric and positive definite, the first term $\mathbf{H}(\omega)$ can be expressed using spectral decomposition as
\begin{equation}
\mathbf{H}(\omega)=\sum_{m=1}^{2n_e+2+S} \frac{\phi_m^T \phi_m}{-(k \omega)^2+2\iu k \omega  \zeta_m \omega_m +\omega_m^2},
\label{eq:Hk}
\end{equation}
where $\phi_m$, $\omega_m$, and $\zeta_m$ are the mode shape, eigenfrequency, and damping ratio of the $m$-th mode of the linear eigenvalue problem in Eq.~\eqref{eq:fem}. Therefore, $\mathbf{H}_{\mathrm{NN}}(\omega)$, $\mathbf{H}_{\mathrm{ex}}(\omega)$, and their derivatives can be obtained analytically from the submatrices of Eq.~\eqref{eq:Hk}. The other derivative of nonlinear forces $\partial \hat{\mathbf{f}}_{\mathrm{n}}\left(\hat{\mathbf{q}}_{\mathrm{N}}, \omega\right)$ will be discussed in the next section with switching-based piezoelectric interface circuits.

\subsection{Electromechanical friction from switching-based interface circuits}

\begin{figure*}[!t]
    \centering
    \includegraphics[width=2\columnwidth,page=3]{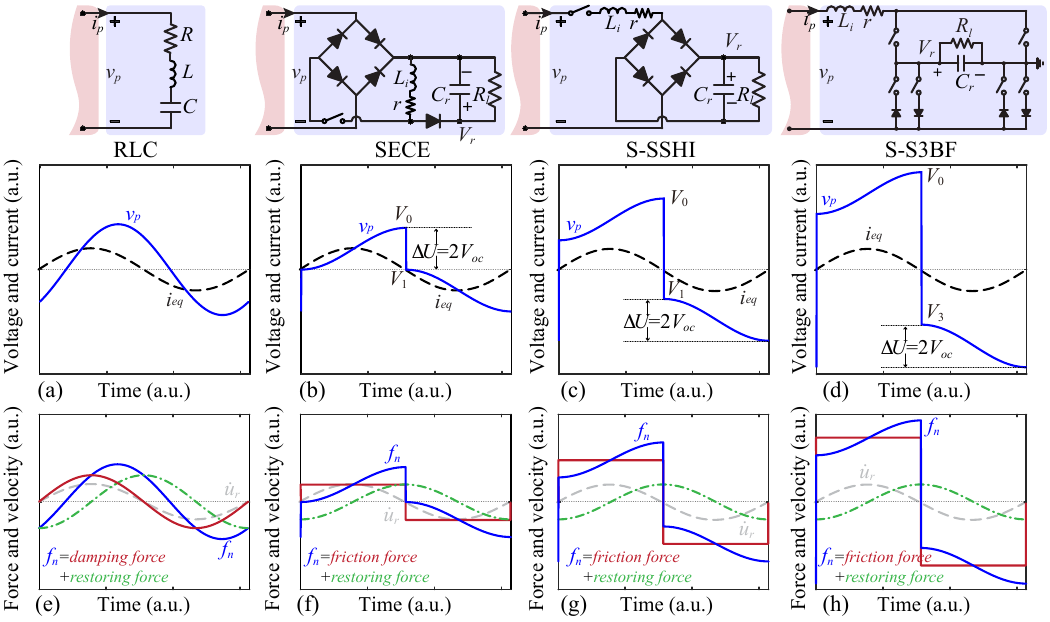}
    \caption{Piezoelectric voltage waveforms and reaction forces induced by different circuits. (a) and (e): Linear electrical load circuit; (b) and (f): SECE interface circuit; (c) and (g): S-SSHI interface circuit; (d) and (h): S-S3BF interface circuit. }
    \label{fig:cir}
\end{figure*}

As shown in Fig.~\ref{fig:resonator}(a), the cantilever-shaped parasitic beams can be modeled as local resonators using their first bending mode, with the piezoelectric reaction force determined by the interface circuit type. The governing equation of the piezoelectric local resonators can be formulated as
\begin{equation}
\begin{aligned}
        {m_r}{{\ddot u}_{r}} + {c_r}{{\dot u}_r} + {k_r}{u_r}+0.5*\alpha_rv_p &=  - {m_r}\left( \ddot{w} + {{\ddot w}_b} \right), \\
        \alpha_r \dot{u}_r & =i_{\mathrm{eq}}=i_p+C_p \dot{v}_p,
\end{aligned}
        \label{eq:parasiticbeam}
\end{equation}
where $\alpha_r v_p$ is the reaction force from the piezoelectric interface circuit, and $\alpha_r \dot{u}_r$ is the equivalent piezoelectric current source proportional to the velocity of the resonator. Since each pair of parasitic beams contains two local resonators, each of which shares half of the total reaction force.
Depending on the choice of the piezoelectric interface circuit, $v_p$ can be altered, leading to linear or nonlinear piezoelectric reaction forces $f_n$ in the mechanical domain. Conventional linear loads such as resistance, capacitance, and inductance shunting for energy harvesting or dynamic tuning maintain linear electromechanical coupling dynamics, whose waveforms are shown in Fig.~\ref{fig:cir}(a). These linear loads can be represented with electrical impedance, and their voltage waveforms $v_p$ have a linear and harmonic shape. 

Unlike linear interface circuits, synchronized switching-based circuits utilize transistors as switches to control the flow of charge \cite{liang2017synchronized}, thus enabling the piezoelectric voltage switching behavior, as shown with the waveforms in Figs.~\ref{fig:cir}(b)--(d).  The phase difference between $v_p$ and $i_{\mathrm{eq}}$ can be eliminated for higher energy harvesting capacity using diodes and switches. Synchronized voltage switching occurs at the current crossing zero point with the help of the $L_i$-$C_p$ resonance, as shown in Figs.~\ref{fig:cir}(b)--(d). During the $LC$ resonance, the piezoelectric voltage $v_p$ flips from $V_0$ to $V_1$ or $V_3$, where the subscript number indicates the flipping times. For SECE in Fig.~\ref{fig:cir}(b), the voltage flip is carried out from $V_0$ to $V_1$ at the voltage falling edge \cite{chen2019revisit}, where $V_1=0$. For S-SSHI in Fig.~\ref{fig:cir}(c) and S-S3BF in Fig.~\ref{fig:cir}(d), they both flip from $V_0$ to negative values $V_1$ and $V_3$ at the voltage falling edge \cite{guyomar2011recent,zhao2020series}. The gradual increase in the piezoelectric voltage and the harvesting capability can be attributed to the increased number of flipping times. 

The classical modeling method for these synchronized switching-based interface circuits is the impedance-based method, which linearizes the piecewise-nonlinear voltage functions. Thus, the interface circuits can be modeled as a series of resistors, inductors, and capacitors shown in Fig.~\ref{fig:cir}(a). The piezoelectric voltage $v_p$ is represented as a first-order harmonic \cite{liang2011impedance,zhao2022graded}.
Although the formulation is concise and straightforward, the nonlinear effect and the nonlinear coupling with the host structure are neglected. This approximation does not adequately represent the effects of amplitude-dependent response and energy transfer to higher harmonics, which are crucial when considering the coupled dynamics of electromechanical metamaterials. Only recent studies have begun to incorporate nonlinearities arising from piezoelectric interface circuits into the modeling of electromechanically coupled dynamics at the resonator level, such as the nonlinearity introduced by a full-bridge rectifier \cite{leadenham2020mechanically}, allowing more accurate predictions of the harvested power and voltage waveforms.

To model the reaction forces of synchronized switching circuits, the formalization of the relationship between piezoelectric voltage $v_p$ and nonlinear effects should be determined. As shown in Figs.~\ref{fig:cir}(b)--(d), except for the bias-flip actions at the current or velocity crossing zero points, the circuits are in an open-circuit state to charge the piezoelectric clamped capacitor $C_p$. Thus, the open circuit voltage $V_{\mathrm{oc}}$ of the piezoelectric patch in Fig.~\ref{fig:cir} can be computed as the voltage accumulation on $C_p$ during a quarter of a vibration cycle, $V_{\mathrm{oc}}$ reads
\begin{equation}
    V_{\mathrm{oc}}=\frac{1}{2C_p} \int_{0}^{\frac{\pi}{\omega}} i_{\mathrm{eq}}(t) \mathrm{d}t=\frac{1}{2C_p} \int_{0}^{\frac{\pi}{\omega}} \alpha_r \dot{u}_r(t) \mathrm{d}t = \frac{\alpha_rU_r}{C_p}.
\end{equation}
Furthermore, $v_p$ can be treated as a general piecewise equation
\begin{equation}
    {v_p}\left(t \right) = \left\{ 
    \begin{aligned}
        & -V_M +\frac{1}{C_p} \int_{0}^{t} i_{\mathrm{eq}}(t) \mathrm{d}t,& 0 \leq  t < \frac{\pi}{\omega};\\
       & V_M +\frac{1}{C_p} \int_{\frac{\pi}{\omega}}^{t} i_{\mathrm{eq}}(t)\mathrm{d}t,  &\frac{\pi}{\omega}  <  t  \leq \frac{2\pi}{\omega},
    \end{aligned}
    \right.
    \label{eq:v_p}
\end{equation}
where $V_M$ is the ending piezoelectric voltage of a given synchronized switching circuit. For SECE, one bias-flip action with the ending voltage $V_1=0$ exists. For S-SSHI or S-S3BF, one or three bias-flip actions exist with the ending voltage $V_1$ or $V_3$ determined by the quality factor and the load conditions of the interface circuits \cite{zhao2020series}. By electromechanical analogy, the piezoelectric reaction force can be formulated as
\begin{equation}
\begin{aligned}
     f_n\left(t \right)&=\alpha_r {v_p}\left(t \right) \\
     &= \left\{ 
    \begin{aligned}
        & -\alpha_r V_M +\frac{\alpha_r^2}{C_p} \left [ u_r(t)-u_r(0) \right ], & 0 \leq  t < \frac{\pi}{\omega};\\
       & \alpha_r V_M +\frac{\alpha_r^2}{C_p}  \left [ u_r(t)-u_r(\frac{\pi}{\omega}) \right ],  &\frac{\pi}{\omega}  <  t  \leq \frac{2\pi}{\omega},
    \end{aligned}
    \right.
\end{aligned}
    \label{eq:f_nl1}
\end{equation}
where $u_r(t)$: $u_r(\hat{u}_{r,H},t)$ is also defined as a Fourier series up to the harmonic order $H$. Since $v_p$ is an odd function, only odd terms of the Fourier series contribute to harmonic generation. Assuming that the first harmonic still dominates the electromechanical dynamics, $u_r(0)$ and $u_r(\frac{\pi}{\omega})$ can be replaced by the magnitude of $u_r$, which means the bias flip occurs at the displacement extrema (i.e., the current and velocity zero-crossings).

Evidently, that $f_n(t)$ is a continuous but non-smooth reaction force that depends on the phase of the relative velocity of the resonator $\dot{u}_r$. This force can be simplified using a sign function as
\begin{equation}
    f_n\left(t \right)=\frac{\alpha_r^2}{C_p} U_{r}\left( 1-\Tilde{V}_M \right) \mathrm{sign} \left (\dot{u}_r \right)+\frac{\alpha_r^2}{C_p}u_r,
    \label{eq:f_nl2}
\end{equation}
where $U_{r}$ is the magnitude of $u_r$, and $\Tilde{V}_M$ is $V_{\mathrm{oc}}$ normalized. This nonlinear reaction force consists of two parts: 
\begin{enumerate}
    \item A linear restoring force corresponds to the open circuit stiffness $\alpha_r^2 / C_p$;
    \item A nonlinear Coulomb type friction force with the coefficient of friction $\frac{\alpha_r^2}{C_p} U_{r}\left( 1-\Tilde{V}_M \right)$ is determined by the displacement amplitude $U_r$ and the ending flipping voltage $\Tilde{V}_M$.
\end{enumerate}

Therefore, the synchronized switching circuits enabled by the piezoelectric resonator in Fig.~\ref{fig:resonator}(a) can be considered equivalent to a nonlinear local resonator in Fig.~\ref{fig:resonator}(b), incorporating an additional restoring force and nonlinear friction from the interface circuits, whose values can be determined by the velocity and displacement of the nonlinear resonator, as shown in Figs.~\ref{fig:resonator}(c) and (d), respectively. As shown in Figs.~\ref{fig:cir}(f)--(h), electromechanical friction can be increased with an increased ending flipping voltage $\Tilde{V}_M$ for stronger nonlinear damping and energy harvesting abilities. 

As a result of the sign function, no analytical derivative of $f_n$ is available for the Jacobian in Eq.~\eqref{eq:dR}, which impedes computational efficiency.  However, observing Eq.~\eqref{eq:f_nl2}  shows that the nonlinear force can be regulated with a hyperbolic tangent function as
\begin{equation}
    f_n\left(t \right)\approx\frac{\alpha_r^2}{C_p} U_{r}\left(  1-\Tilde{V}_M \right) \tanh \left (\beta \dot{u}_r \right)+\frac{\alpha_r^2}{C_p}u_r,
    \label{eq:f_nl3}
\end{equation}
where $\beta$ is the regularization factor (see Appendix \ref{appendixC} for details). When $\beta$ is large enough, the hyperbolic tangent function recovers the sign function in Eq.~\eqref{eq:f_nl2} with an analytical derivation. The expression of $f_n$ involves  displacement $u_r$ and velocity $\dot{u}_r$ of the local resonator; thus, its total derivative is given as
\begin{equation}
    \mathrm{d} f_n=\frac{\alpha_r^2}{C_p} U_{r}\left( 1-\Tilde{V}_M \right)\beta \text{sech}^2 \left (\beta \dot{u}_r \right) \mathrm{d} \dot{u}_r+\frac{\alpha_r^2}{C_p} \mathrm{d} u_r,
        \label{eq:df_nl}
\end{equation}
where the sech function represents the hyperbolic secant function. 

\begin{figure*}[!t]
    \centering
    \includegraphics[width=2\columnwidth,page=4]{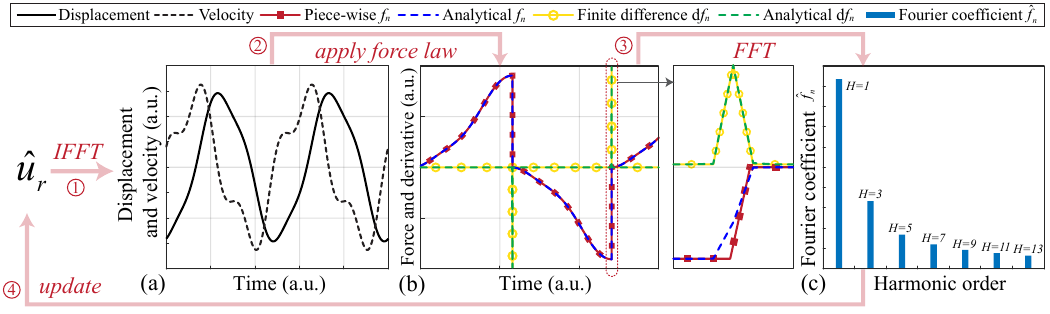}
    \caption{Procedures of the alternating frequency and time. (a) IFFT of the ansatz of the solution; (b) Time domain nonlinear reaction force and its derivative; (c) FFT of the time domain reaction force.  }
    \label{fig:aft}
\end{figure*}

As shown in Eq.~\eqref{eq:f_nl3}, the nonlinearity of the reaction force mainly originates from the hyperbolic tangent function associated with the velocity $\dot{u}_r$ due to the switching behavior. 
With the above derivation of the analytical expressions of the reaction force and its derivative of the interface circuit, the calculation of the Fourier coefficients of $\hat{f}_n$ and $\mathrm{d}\hat{f}_n$ can be resolved using the alternating frequency time (AFT) method \cite{Cameron1989} in Fig.~\ref{fig:aft}. At a given frequency $\omega$, an initial estimate of the unknown $\hat{u}_r$ can be determined using the numerical continuation technique. With the SECE interface circuit taken as an example, the implementation of the AFT can be divided into four steps:
\begin{enumerate}
    \item  An inverse fast Fourier transform (IFFT) is performed to obtain the displacement and velocity in the time domain, with the initial guess as shown in Fig.~\ref{fig:aft}(a).
    \item The force law is applied according to the chosen interface circuit to calculate the reaction force in the time domain and its derivative as shown in Fig.~\ref{fig:aft}(b).
    \item Fast Fourier transform (FFT) is performed to recover the frequency domain Fourier coefficients of $\hat{f}_n$ and $\mathrm{d}\hat{f}_n$ as shown in Fig.~\ref{fig:aft}(c).
    \item The unknown $\hat{u}_r$ is updated by minimizing the residual function in Eq.~\eqref{eq:condense2} using numerical methods, i.e., the Newton--Raphson method.
\end{enumerate}

As shown in Fig.~\ref{fig:aft}(b), choosing a relatively large $\beta$ ensures that the waveforms of the analytical $f_n(t)$ and its derivative $\mathrm{d}f_n(t)$ in Eqs.~\eqref{eq:f_nl3} and \eqref{eq:df_nl} agree well with those of the piecewise $f_n(t)$ in Eq.~\eqref{eq:f_nl1} and its derivative using finite difference, demonstrating the effectiveness of the nonlinear reaction force regulated for faster convergence of solutions. The amplitudes of $\hat{f}_n$ for different harmonic orders $H$ are shown in Fig.~\ref{fig:aft}(c). In addition to the first-order harmonic that represents the impedance in other documentation \cite{zhao2023circuit,liang2011impedance}, multiple odd harmonics occur due to the nonlinear switching behavior. 

In fact, the conventionally utilized first-order impedance-based model is a simplified case of the nonlinear reaction force in Eq.~\eqref{eq:f_nl3}. With the assumption of the first-order harmonic case and large $\beta$, $\tanh \left (\beta \dot{u}_r \right)$ can be replaced with its leading harmonic; therefore, the reaction force is simplified as
\begin{equation}
\begin{aligned}
        f_n^{H=1}\left(t \right)&=\frac{\alpha_r^2}{C_p} U_{r}\left( 1-\Tilde{V}_M \right) \frac{4\dot{u}_r}{\pi\dot{U}_r}  +\frac{\alpha_r^2}{C_p}u_r\\
       &= \frac{4\alpha_r^2}{C_p \pi\omega}  \left(1-\Tilde{V}_M\right)\dot{u}_r+\frac{\alpha_r^2}{C_p}u_r,
\end{aligned}
    \label{eq:f_nl3_h1}
\end{equation}
where  $\dot{U}_r$ is the amplitude of the velocity $\dot{u}_r$.  $ f_n^{H=1}$ introduces the same electrically induced damping and stiffness as stated in previous articles \cite{chen2019revisit}. Tuning the load conditions or switching logic in the S-SSHI or S-S3BF interface circuits ensures that $V_M$ can be tuned for  positive damping ($1-\Tilde{V}_M>0$) and negative damping ($1-\Tilde{V}_M<0$) \cite{zhao2023jump}. For simplicity of analysis, we employ heavy ($R_l=0$) and zero load ($R_l=\infty$) conditions for the S-SSHI and S-S3BF interface circuits in this paper. Under these conditions \cite{zhao2020series},  $\Tilde{V}_M$ will be $2\gamma/(1+\gamma)$ for the S-SSHI circuit and $(4\gamma-2)/(1+\gamma)$ for S-S3BF, where $\gamma$ is the flip factor associated with the quality factor of the bias-flip actions \cite{liang2011impedance}.

\subsection{Nonlinear band structures and frequency responses}
\label{sec:dispersion&frf}
Building on the general formulation of various synchronized switching-based interface circuits, we investigate their influence on wave propagation through the analysis of nonlinear band structures and frequency response functions. These analyzes are conducted for infinitely periodic and finite metamaterial systems, with the appropriate boundary conditions applied in each case.

\begin{figure}[!t]
    \centering
    \includegraphics[width=1\columnwidth,page=5]{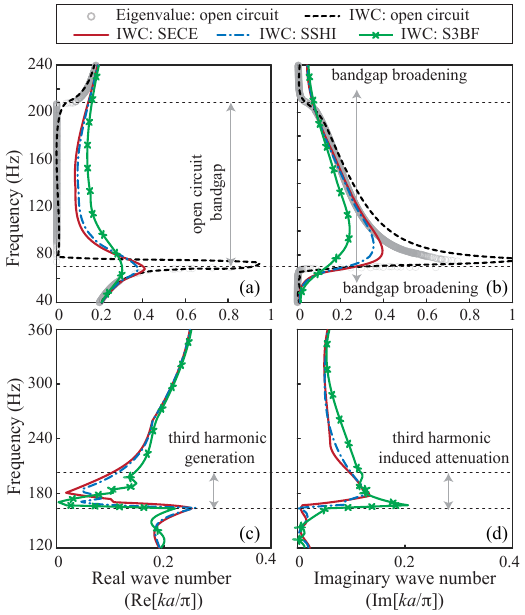}
    \caption{Nonlinear band structures of the metamaterial beam. (a) and (b): Real and imaginary wave numbers for the first-order harmonic wave propagation; (c) and (d): Real and imaginary wave numbers for the third harmonic wave propagation.}
    \label{fig:dispersion}
\end{figure}

The key distinction between linear dispersion relations and nonlinear band structures lies in the dependence on the input amplitude \cite{fang2024advances}. While linear dispersion assumes amplitude-independent behavior, nonlinear systems exhibit amplitude-dependent dynamics. These nonlinearities give rise to energy transfer between harmonics, resulting in the emergence of higher-order harmonic components in the wave field \cite{fang2018wave, zhao2024nonlinear}, rendering the traditional linear dispersion concept insufficient.
For reference, the linear dispersion relationship under the open circuit condition can be obtained for an infinitely long metamaterial beam, formulated using the finite element approach introduced in Sec.~\ref{sec:rom} with the single-element unit cell documented in Appendix \ref{appendixA}. The unknowns of node $n$ and its adjacent nodes can be assumed to follow Bloch's theorem as
\begin{equation}
\begin{aligned}
        &\mathbf{q}_{n}=\left[W{\rm e}^{\iu(kx-\omega t)},\Theta{\rm e}^{\iu(kx-\omega t)}, U_r{\rm e}^{\iu(kx-\omega t)}\right]^\transpose, \\
         &\mathbf{q}_{n}={\rm e}^{\iu k a}\mathbf{q}_{n-1}={\rm e}^{-\iu k a}\mathbf{q}_{n+1},
\end{aligned}
\end{equation}
where $k$ represents the wave number, and ${\rm e}^{\pm \iu k a}$ is the spatial phase shift of the amplitudes $W$, $\Theta$, and $U_r$.  By selecting the repeating pattern in the assembly of the system, the linear eigenvalue problem of the metamaterial beam can be expressed as
\begin{equation}
    \left(\mathbf{K}_{\mathrm{3er}}-\omega^2\mathbf{M}_{\mathrm{3er}}+\iu\omega\mathbf{C}_{\mathrm{3er}}\right)\mathbf{q}_{n}=0,
\end{equation}
where $\mathbf{M}_{\mathrm{3er}}$, $\mathbf{K}_{\mathrm{3er}}$, and $\mathbf{C}_{\mathrm{3er}}$ are the repeating matrices in the assemblies of the mass, stiffness, and damping matrices of the metamaterial beam with three single-element unit cells. The open circuit condition only introduces electrically induced stiffness due to the piezoelectric clamp capacitor $C_p$. Therefore, the Newton method can solve this linear eigenvalue problem at each frequency for a complex wave number. The solutions of the real and imaginary wave numbers are indicated with gray circles in Figs.~\ref{fig:dispersion}(a) and (b). 

When nonlinearities are present, the common procedure is to apply linearization of the nonlinear reaction force and obtain the proportional forces of the velocity and displacement, i.e., the first-order harmonic, as shown in Eq.~\eqref{eq:f_nl3_h1}. The linearized effect of interface circuits on wave propagation can be analyzed by substituting their corresponding electrically induced damping and stiffness into local resonators \cite{sugino2018design,bao2020manipulating,jian2023analytical,bao2025piezoelectric}. However, the higher-order energy exchanges and their effects due to harmonic generation are neglected. To include these effects, recent efforts feature harmonic balance-based methods \cite{fang2018wave,wang2024nonlinear} and perturbation-based multiple scales methods \cite{bae2020amplitude,bukhari2020spectro,fortunati2022nonlinear,zhao2024wideband,shen2024wave}. While most of them require multiple harmonics or scale expansions of the analytical nonlinear reaction force, this paper proposes an alternative way to resort to the solution of the wave field by the ROM-HB method in Sec.~\ref{sec:rom}. The wave number of the nonlinear band structure can be estimated as the value of the wave number $k$ that maximizes the correlation between the computed wave field and an assumed field at frequency $\omega_0$
\begin{equation}
\mathcal{C}\left(k\right)=
\frac{\left|\int w\left(x,\omega_0\right) \cdot v^*\left(x, k\right) \mathrm{d} x\right|}{\sqrt{\int\left|w\left(x, \omega_0\right)\right|^2 \mathrm{d} x \cdot \int\left|v\left(x, k\right)\right|^2 \mathrm{d} x}},
\end{equation}
where $*$ denotes the complex conjugate. $v=V{\rm e}^{\iu(kx-\omega_0 t)}$ is the assumed solution of the wave field. This method is referred to as the inhomogeneous wave correlation method \cite{zhao2022graded,van2018measuring}. The IWC function $\mathcal{C}$ typically has a well-defined maximum at which the two transformations exhibit the highest correlation, allowing for an estimation of the complex wave number from
\begin{equation}
    k=\mathrm{agr} \, \underset{k}{\mathrm{max}} \mathcal{C}\left(k\right).
\end{equation}

To mimic an infinitely long metamaterial beam with nonlinear local resonators, we maintain the same unit cell length $a$ but increase the number of unit cells to 20, covering a 0.45 m section of a 1 m cantilevered metamaterial beam. Other parameters remain the same as shown in Table \ref{tab:para}.  
By substituting the solutions of the wave field into the 20 pairs of nonlinear local resonators and ensuring that the frequency corresponds to different harmonics, multiple nonlinear band structures for different harmonics are shown in Fig.~\ref{fig:dispersion}. Evidently, the bandgap ranges under open-circuit conditions obtained by eigenvalue analysis and the IWC method generally agree with each other. The slight deviation in Fig.~\ref{fig:dispersion}(b) might be attributed to the finite number of nonlinear resonators applied in the FEM modeling. The first-order harmonic wave numbers with different synchronized switching-based interface circuits are shown in Figs.~\ref{fig:dispersion}(a) and (b). Compared with the open-circuit case, the bandgaps become broader with these interface circuits. The S-S3BF interface circuit yields the widest bandgap due to its stronger electromechanical friction. The third-order harmonic wave numbers are also presented in Figs.~\ref{fig:dispersion}(c) and (d). As a result of harmonic generation, we can also observe another attenuation range approximately three times the original resonant frequency of the nonlinear local resonators, which demonstrates the energy exchange in nonlinear settings. 

\begin{figure*}[!t]
    \centering
    \includegraphics[width=2\columnwidth,page=6]{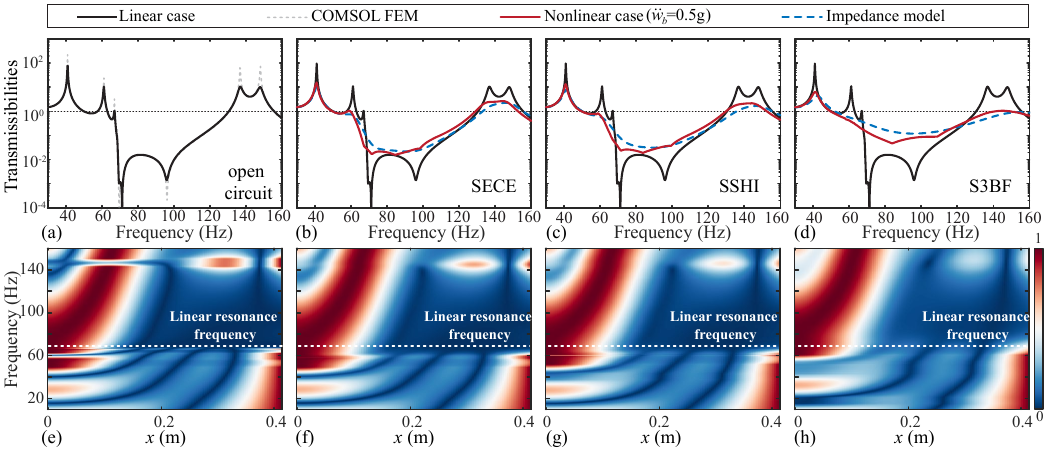}
    \caption{Nonlinear bandgap and spatial frequency analyses of the metamaterial beam. (a) and (e): Open circuit linear case; (b) and (f): Nonlinear case with SECE interface circuit; (c) and (g): Nonlinear case with S-SSHI interface circuit; (d) and (h): Nonlinear case with S-S3BF interface circuit. The color bar indicates the normalized amplitude of the first-order harmonic displacement across the host beam. }
    \label{fig:bandgap}
\end{figure*}

\begin{figure}[!t]
    \centering
    \includegraphics[width=1\columnwidth,page=7]{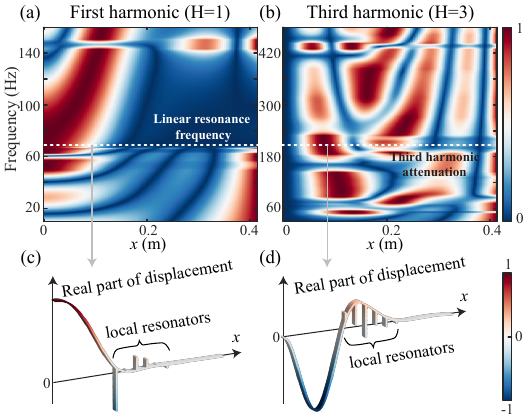}
    \caption{Spatial analyses and real part displacement fields of the first- and third-order harmonic wave propagation along the metamaterial beam with SECE interface circuits. (a) and (c): First-order harmonic wave propagation; (b) and (d): Third-order harmonic wave propagation. The displacement fields in (c) and (d) are normalized by their maximum displacement amplitude. }
    \label{fig:thirdharmonic}
\end{figure}

While nonlinear band structures provide insight into wave propagation in infinitely periodic systems, the behavior can differ significantly in finite metamaterials due to boundary effects and modal interactions. Here, we discuss the nonlinear frequency responses of a clamped-free nonlinear metamaterial beam with six pairs of nonlinear local resonators. Two piezoelectric patches are attached to both sides of a parasitic beam, and the four patches of each pair of nonlinear local resonators are connected in series to boost the input voltage for the interface circuit. The host beam is 0.4125 m long and is discretized into 132 beam elements, allowing local resonators to be attached to the nodal points of the host beam. 
Table \ref{tab:para} lists the detailed parameters obtained from the experiments. The system is first condensed into six pairs of nonlinear local resonators and solved using the ROM-HB method, through which the solutions of the host beam can be recovered.

The customized linear FEM model is first compared against the COMSOL model in Fig.~\ref{fig:bandgap}(a). The COMSOL model consists of a host beam, modeled by the plane stress solid mechanics module, and local resonators, modeled by the truss module with assigned lumped parameters. The length of the host beam is much larger than its thickness; thus, the plane stress model can efficiently approximate an Euler–Bernoulli beam without shear deformation \cite{timoshenko2012theory}. The host beam is discretized along its span with a target element length of 1\% of the unit cell length and is mapped to five layers through the thickness. Each local resonator is meshed by a single truss element, kinematically tied to the host beam at its root position \cite{pu2024multiple}. The tip transmissibility of the metamaterial beam under open-circuit conditions, computed using both methods, shows excellent agreement, thereby validating the accuracy and reliability of the underlying linear system model. By introducing a base excitation with an acceleration $\ddot{w}_b=0.5$ g, we employed $H=9$ harmonic orders to compute the nonlinear frequency responses of the metamaterial beam. The transmissibility is determined as the ratio between the maximum displacement amplitude of the beam tip point, considering nine harmonics, and the base excitation displacement. The nonlinear frequency responses with the SECE, S-SSHI, and S-S3BF interface circuits are shown in Figs.~\ref{fig:bandgap}(b)--(d). From SECE to S-S3BF, the bandgap becomes broader, demonstrating stronger friction and energy harvesting abilities. In addition to the broader bandgap, the modal frequencies are attenuated due to the nonlinear friction and modal coupling capabilities of the nonlinear local resonators. This process can also be observed from the spatial frequency graphs along the host beam in Figs.~\ref{fig:bandgap}(e)--(h), which show the normalized first-order harmonic amplitude of the wave field. The initially sharp transitions near the linear resonant and modal frequencies become increasingly blurred due to the nonlinear electromechanical friction imposed by the switching-based interface circuits.

The third harmonic generation of the metamaterial beam with the SECE interface circuit is shown in Fig.~\ref{fig:thirdharmonic}. As a result of electromechanical friction, the third harmonic can be generated and propagated along the host beam. In addition to the first-order harmonic-induced bandgap, the collective behaviors of the higher harmonics also create bandgaps at higher harmonic frequencies. As shown in Fig.~\ref{fig:thirdharmonic}(b), around three times the linear resonant frequency, an attenuation range exists, indicated by dark blue for the third-harmonic wave propagation along the host beam, which confirms the third-harmonic band structures in Figs.~\ref{fig:dispersion}(c) and (d). Closely inspecting the real part of the displacement field in Figs.~\ref{fig:thirdharmonic}(c) and (d) shows that the third-order harmonic-induced attenuation can be attributed to the out-of-phase movements between the nonlinear local resonators and the host beam. Notably, the higher-order harmonic-induced attenuation, i.e., the third-order harmonic-induced attenuation, depends on the first harmonic vibration of the nonlinear local resonators.

\begin{figure}[!t]
    \centering
    \includegraphics[width=1\columnwidth,page=8]{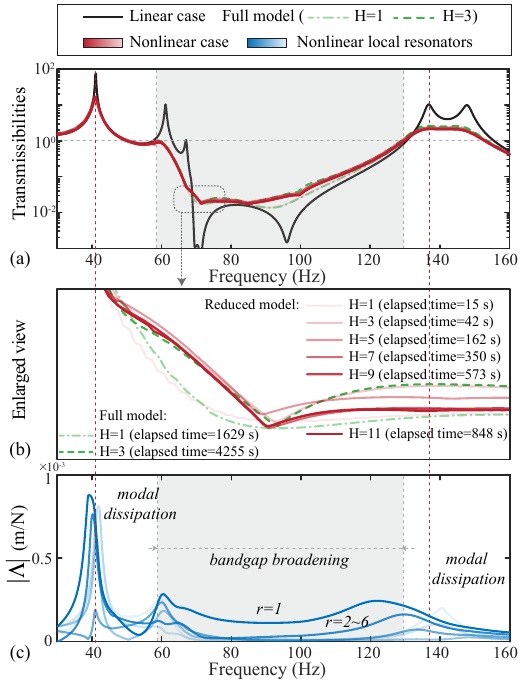}
    \caption{(a) Transmissibilities of the metamaterial beam calculated with the ROM-HB method and full-scale modeling method; (b) Enlarged view and computational effort; (c) Frequency response function of different nonlinear local resonators. }
    \label{fig:converge}
\end{figure}

Through the same solver settings over the frequency range from 30--160 Hz, the convergence and computational effort of the proposed ROM-HB method and the full-scale method are shown in Fig.~\ref{fig:converge}. Increasing the harmonic order causes the solutions of the ROM-HB method to gradually converge to those of higher harmonic orders. To ensure the adequacy of the truncation order, we conducted a convergence check by varying the harmonic order up to $H=11$. After $H=7$, the nonlinear bandgap curve converges rapidly with negligible differences. Beyond $H=9$, higher-order Fourier coefficients of the resonator displacements are less than 1\% compared with their fundamental Fourier coefficients, confirming that $H=9$ truncation is sufficient to capture the essential higher-harmonic contributions and avoid computational overhead. Compared with the full-scale model, the ROM-HB method not only achieves the same accuracy but also increases computational efficiency by around 100 times \footnote{Computed on a MacBook Pro with an Apple M1 Pro chip and 32 GB of RAM.}, which facilitates the computation of nonlinear coupling effects. 
As shown in Fig.~\ref{fig:converge}(c), the amplitude of the \textit{frequency response function} $|\Lambda|$ is defined as the ratio of the maximum displacement amplitude of each local resonator $u_r$ to its root position displacement of the host beam $w$. The presence of nonzero $|\Lambda|$ values outside the linear resonant frequency range indicates strong coupling between the host beam and the nonlinear local resonators, highlighting the metamaterial’s capability for broadband vibration attenuation.

\begin{figure}[!t]
    \centering
    \includegraphics[width=1\columnwidth,page=9]{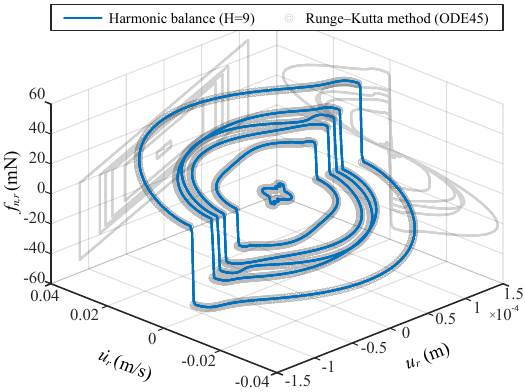}
    \caption{Limit cycles of nonlinear local resonators with their nonlinear reaction forces at 40 Hz.}
    \label{fig:cyc}
\end{figure}

In addition to the frequency domain method, the system can also be solved using the Runge--Kutta time domain integration method. Transforming the governing equation Eq.~\eqref{eq:fem} into a state-space formulation enables the steady-state solution to be reached after a sufficiently long computation time with forced external excitation \cite{zhao2024nonlinear}. Fig.~\ref{fig:cyc} shows the agreement between the recovered time domain limit cycles of the nonlinear resonators at 40 Hz with the IFFT of the solutions from the ROM-HB method and the results from the direct time domain integration. As a result of electromechanical friction, the projections of the limit cycles on the velocity-force plane are shifted with respect to the sign of velocity, which provides possibilities for higher harmonic generation. 

\section{Experimental Realization}

This section presents the verification and demonstration of the proposed electromechanical nonlinear metamaterial. From the mechanical perspective, the influence of synchronized switching-based interface circuits on elastic wave propagation is experimentally verified, particularly the electromechanical friction that enables the broader bandgap range and higher harmonic-induced vibration attenuation. From an electrical standpoint, we evaluate the power performance of the system under varying excitation amplitudes, as well as the power consumption requirements for IoT applications, specifically those involving temperature and acceleration sensing. These results demonstrate the effectiveness of synchronized switching-based interface circuits as AC–DC converters for self-powered sensing. Although S-SSHI and S-S3BF interface circuits have demonstrated stronger electromechanical damping effects in previous studies \cite{zhao2023circuit,zhao2020series} and friction abilities for broader bandgap ranges, as shown in Fig.~\ref{fig:bandgap}, they require more electrical components and additional gate drivers for synchronized switching sequence control. To verify the electromechanical friction effect induced via synchronized switching, we employed the SECE interface circuit as a preliminary example in all experiments. More advanced interface circuits will be implemented in further studies. 
For the nonlinear bandgap, harmonic generation, and power performance in Figs.~\ref{fig:exbandgap},~\ref{fig:exthird}, and~\ref {fig:power}, we employed six SECE interface circuits for each pair of parasitic beams to store the harvested energy in the same capacitor $C_r$. Moreover, for the self-powered sensing demonstrated in Fig.~\ref{fig:iotpower}, we connect the first pair of parasitic beams to a compact IoT board that combines the SECE interface circuit, an energy management circuit, and a low-power Bluetooth beacon.

\subsection{Experimental setup}
\begin{figure*}[!t]
    \centering
    \includegraphics[width=2\columnwidth,page=10]{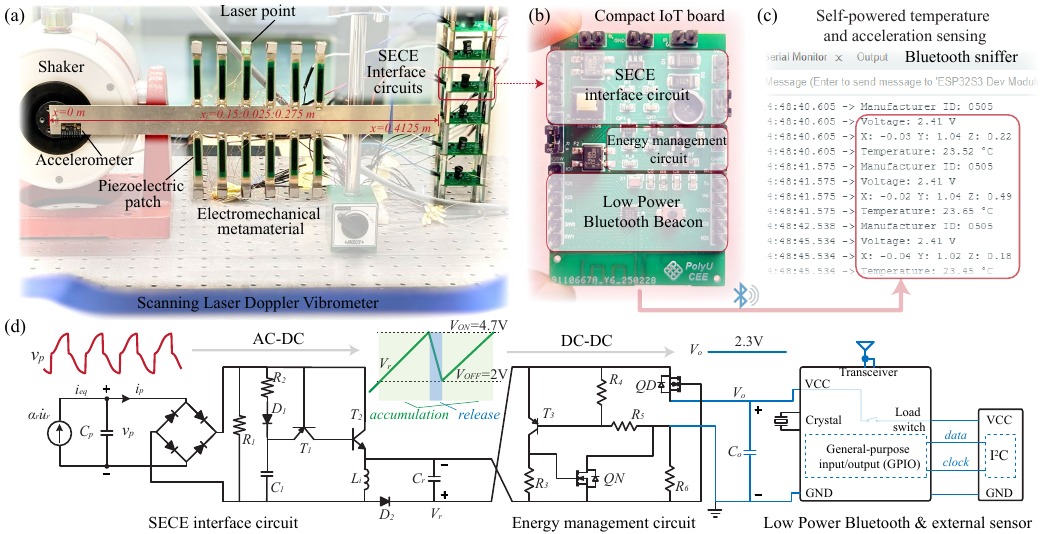}
    \caption{Experimental setup of the proposed electromechanical metamaterial. (a) Clamped-free piezoelectric metamaterial beam with 6 SECE interface circuits; (b) Compact IoT board includes the SECE interface circuit, energy management circuit, and a low-power Bluetooth beacon; (c) Screenshot of self-powered temperature and acceleration sensing captured by a remote Bluetooth sniffer (ESP32-S3); (d) Schematic of the compact IoT board. }
    \label{fig:setup}
\end{figure*}

\begin{table}[t!]
     \footnotesize
    \caption {Experimental parameters} \label{tab:para}
    \begin{center}
    \setlength{\tabcolsep}{1mm}{
    \begin{tabular}{llll}
    \toprule
    \multicolumn{4}{c}{\textbf{Metamaterial Node}}     \\
    \midrule
    \multicolumn{4}{l}{\textbf{Host Beam}}                                \\
    Size   & 412.5$\times$30$\times$1 (mm$^3$) & Material  & Aluminum   \\
    \multicolumn{4}{l}{\textbf{Piezoelectric Local Resonator}}                \\
     Size  & 65$\times$10$\times$1 (mm$^3$) & Material  & Aluminum  \\
     Tip mass  & 10$\times$10$\times$10 (mm$^3$) & Material  & Steel  \\
     $\alpha_r$   &  0.48 (mN/V)   & $C_p$ & 5 (nF) \\ 
     $\omega_r$ & 2$\pi\times$69.4 (rad/s) & \multicolumn{2}{l}{$x_j=150:25:275$  (mm) }  \\
    \multicolumn{4}{l}{\textbf{SECE Interface Circuit}}                \\
    Diodes  & SS16   & $L_i$    & 10 (mH)        \\
    $T_1$\&$T_2$  & ZXTD4591  & $C_r$    & 47 ($\mu$F)    \\
    Rectifier & MB6S   &  $\gamma$ &   -0.33\\
    \multicolumn{4}{l}{\textbf{Energy Management Circuit}}        \\
    QN & BSS123N   & $T_3$   & BC856B         \\
    QD & BSS159N   \\
    \multicolumn{4}{l}{\textbf{Bluetooth beacon}}       \\         
    $C_o$ & 10 ($\mu$F)  &  Chip  & Inplay IN100         \\
    Cystal   & 26 (MHz) & Accelerometer & BMA400    \\
    \bottomrule
    \end{tabular}}
    \end{center}
\end{table}

As shown in Fig.~\ref{fig:setup}, the experimental prototype comprises the metamaterial beam, SECE interface circuits, and a compact version of the IoT board, which collectively form the proposed EMetaNode. The metamaterial beam, with six parasitic beams on each side, is fabricated by water-jet cutting from a 1 mm aluminum plate. Each parasitic beam is attached with two piezoelectric patches on both sides and one tip mass. The four patches of one pair of parasitic beams are connected in series to boost the piezoelectric voltage for one piece of the six SECE interface circuits on the right side of Fig.~\ref{fig:setup}(a). Through wave propagation and power performance measurements, all SECE interface circuits are connected to the six pairs of local resonators, introducing electromechanical friction for vibration attenuation and energy conversion to enable self-powered sensing. The applied SECE interface circuit is realized by self-powered switching with an envelope detector ($R_2$, $D_1$, and $C_1$) and a transistor switching path ($T_1$ and $T_2$) in Fig.~\ref{fig:setup}(d), the details of which can be found in \cite{zhao2022graded}. 

In addition to the SECE interface circuit, the compact IoT board (Figs.~\ref{fig:setup}(b) and (d)) includes an energy management circuit and a low-power Bluetooth system on chip (SoC) for DC–DC regulation and self-powered sensing and transmission. The energy management circuit proposed by Teng et al. \cite{teng2023three} features a three-transistor design with adjustable regulation voltage thresholds $V_{\mathrm{ON}}$ and $V_{\mathrm{OFF}}$. The regulated output voltage, $V_o\approx$2.3 V, leverages the gate-threshold voltage of a depletion-mode n-channel MOSFET, $QD$, which is suitable for most IoT SoCs. An ultra-low-power Bluetooth beacon SoC, IN100 \cite{IN100}, is integrated on the IoT board to detect and transmit on-chip temperature and external 3-axis acceleration data (via the BMA400 accelerometer) to remote receivers through continuous Bluetooth broadcasting on channels 37 to 39 with a 1-second interval. As shown in Fig.~\ref{fig:setup}(d), after the SECE interface circuit, the harvested power is accumulated in the storage capacitor $C_r$ until the voltage $V_r$ reaches the threshold voltage $V_{\mathrm{ON}}$ of the energy management circuit. Through this regulation circuit, DC energy is released and regulated to the output voltage $V_o$ to power the Bluetooth beacon. Connecting the external low-power sensor to the load switch of the Bluetooth beacon enables the sensor to also be powered up with $V_o$ when the Bluetooth beacon wakes up from sleep mode. Self-powered sensing and transmission will operate whenever the Bluetooth beacon wakes up. Between the 1-second broadcasting intervals, the Bluetooth beacon and the external sensor will enter sleep mode to save power. Adjusting the onboard selections of $R_5$ achieves different threshold voltages $V_{\mathrm{ON}}$, providing capabilities for heavy-sensing and transmission tasks. 
The parameters of the EMetaNode are shown in Tab.~\ref{tab:para}. The printed circuit board fabrication files, schematics, and configuration codes for this compact IoT board are open-source. They can be accessed at \cite{iotboard}.

\subsection{Experimental results}

We first measure the velocity field of the electromechanical metamaterial beam under different excitation levels. One end of the metamaterial beam is clamped to a shaker (LDS-V406) for base excitation, forming a clamped-free boundary condition. A Polytec laser Doppler vibrometer records the out-of-plane time-domain velocity field along the host beam through repeated acquisitions. The nonlinear coupling effects with the SECE interface circuits are observed by applying a slow sweep signal ranging from 10 to 210 Hz to the clamped side of the nonlinear metamaterial beam at constant acceleration levels. The transmissibilities of the metamaterial beam are defined as the ratio of the frequency responses of the tip velocity to those of the clamped side.  

\begin{figure}[!t]
    \centering
    \includegraphics[width=1\columnwidth,page=11]{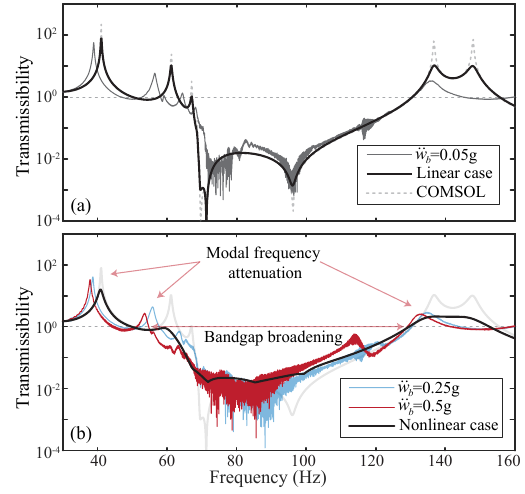}
    \caption{Experimental and theoretical transmissibilities of the nonlinear metamaterial under different base excitation. (a) Linear case with open circuit condition; (b) Nonlinear case with SECE interface circuits.}
    \label{fig:exbandgap}
\end{figure}

\begin{figure}[!t]
    \centering
    \includegraphics[width=1\columnwidth,page=12]{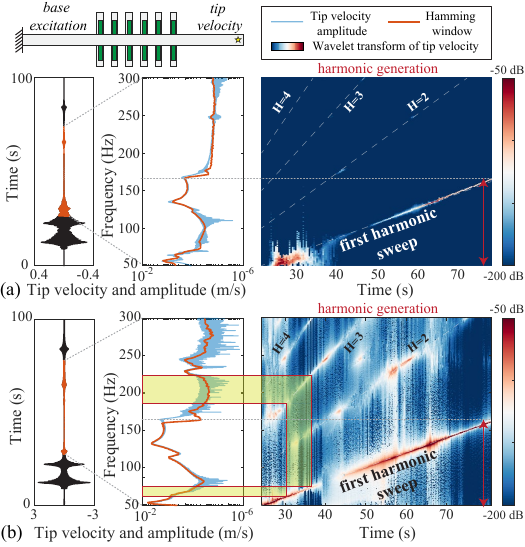}
    \caption{Experimental harmonic generation and higher-order harmonics-induced attenuation of the host beam. (a) Linear case with no harmonic generation; (b) Nonlinear case with higher-order harmonic generation and harmonics-induced vibration attenuation.}
    \label{fig:exthird}
\end{figure}

\begin{figure*}[!t]
    \centering
    \includegraphics[width=2\columnwidth,page=13]{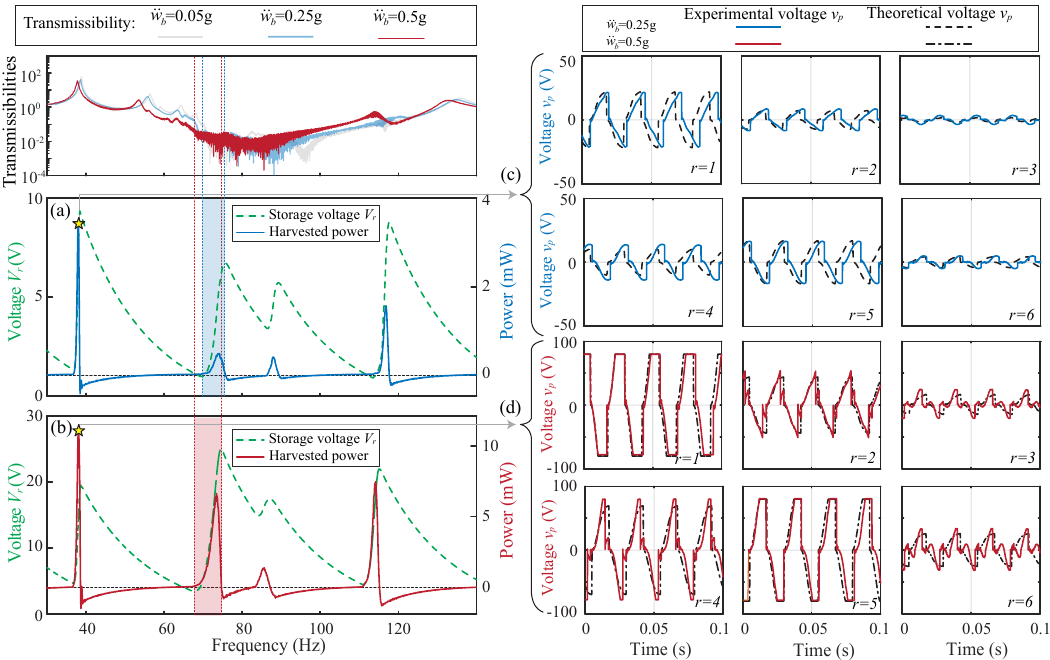}
    \caption{Experimental power performance and piezoelectric voltage waveforms of each pair of parasitic beams. (a) and (c): Power performance of the metamaterial system and the piezoelectric voltage waveforms under $\ddot{w}_b=0.25$ g;  (b) and (d): Power performance of the metamaterial system and the piezoelectric voltage waveforms under $\ddot{w}_b=0.5$ g.}
    \label{fig:power}
\end{figure*}

\begin{figure}[!t]
    \centering
    \includegraphics[width=1\columnwidth,page=14]{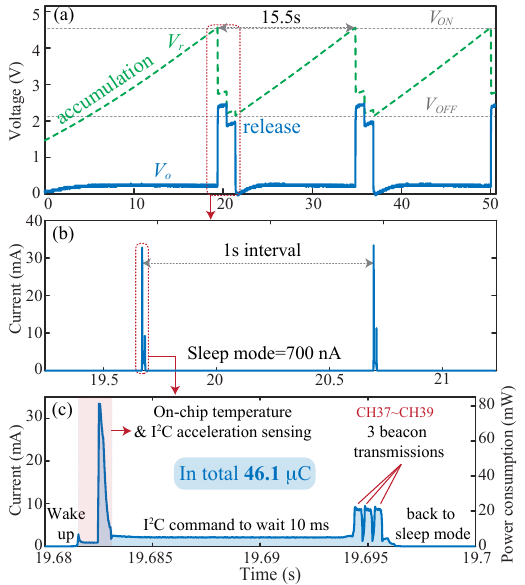}
    \caption{Experimental energy consumption of self-sensing functions. (a) Storage DC voltage $V_r$ of capacitor $C_r$ and the regulated voltage $V_o$ of capacitor $C_o$; (b) Current consumption of sensing and Bluetooth transmission with a 1-second interval; (c) Enlarged view of the current and power consumption for each sensing and transmission process. }
    \label{fig:iotpower}
\end{figure}

\begin{figure}[!t]
    \centering
    \includegraphics[width=1\columnwidth,page=15]{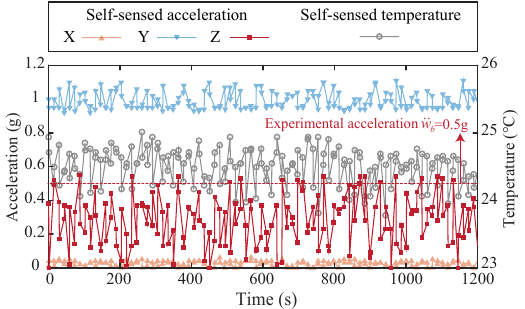}
    \caption{Experimental self-powered sensing of the 3-axis acceleration and temperature data over 1200 seconds under base excitation $\ddot{w}_b=0.5$ g. }
    \label{fig:sensing}
\end{figure}

Through a small base excitation acceleration $\ddot{w}_b=0.05$ g, the tip transmissibility of the metamaterial beam is shown in Fig.~\ref{fig:exbandgap}(a), where the SECE circuits are disconnected to create open circuit conditions as the linear case of the metamaterial beam. The experimental bandgap generally agrees with the theoretical results of the finite element model and COMSOL. The slight differences in the modal frequencies are due to the single-degree-of-freedom assumption of the parasitic beams. The nonlinear transmissibilities of the metamaterial beam, achieved by connecting the SECE interface circuits and increasing base excitation accelerations, are shown in Fig.~\ref{fig:exbandgap}(b). With increases in the base excitation, the nonlinear friction effect of the SECE interface circuits becomes more prominent. This electromechanical friction broadens the bandgap range and enables stronger modal frequency dissipation abilities under larger base excitations. The experimental results agree with the theoretical nonlinear frequency responses discussed in Sec. \ref{sec:dispersion&frf}. 
With increasing base excitation acceleration, the piezoelectric voltage $v_p$ may exceed the combined value of the storage voltage $V_r$ across capacitor $C_r$ and the collector–emitter breakdown voltage $V_{\mathrm{CE}}$ (60 V for $T_3$) of the transistor $T_3$ shown in Fig.~\ref{fig:setup}(d). When this occurs, transistor $T_3$ enters its soft breakdown region, causing the voltage across its collector and emitter to be clamped at $V_{\mathrm{CE}}$ \cite{erickson2007fundamentals}. Therefore, an experimentally determined voltage limit of $v_p\approx$80 V is applied in the modeling of the nonlinear reaction force in Eq.~\eqref{eq:f_nl3} as a saturation, beyond which the nonlinear reaction force is clamped. This effect of force saturation further leads to the softening of the nonlinear local resonators under large excitation accelerations around their resonance and modal frequencies, as shown with the red line in Fig.~\ref{fig:exbandgap}(b).

As a result of the implementation of the fundamental harmonic sweep excitations from the clamped side of the host beam, the harmonic generation in nonlinear cases occurs in the region of nonlinear local resonators and propagates across the host beam. This condition makes the conventional definition of transmissibility between input and output points inappropriate. Thus, to validate the higher-order harmonics-induced attenuation, we compare the wavelet transform of the tip velocity of the metamaterial beam under base excitation $\ddot{w}_b=0.05$ g and $\ddot{w}_b=0.5$ g in Figs.~\ref{fig:exthird}(a) and (b), respectively. The wavelet transforms are calculated using the time-domain signals corresponding to the frequency sweep from 50 Hz to 170 Hz, thereby avoiding the third-harmonic range of the nonlinear local resonators. Indicated with the amplitude of the frequency domain of the tip velocity in Fig.~\ref{fig:exthird}(a), the linear band gap is clearly observed from the linear resonant frequency of the local resonators. Through its wavelet transform, the main harmonic observed is the first-order harmonic sweep from the excitation source. Over 170 Hz, the frequency domain amplitude and the wavelet transform demonstrate no harmonic generation. Unlike the linear case, the nonlinear case with the SECE interface circuits reveals strong harmonic generation around the resonant frequency of the nonlinear local resonators across the host beam, which introduces the third-order harmonic-induced attenuation as shown with the yellow region in the frequency domain amplitude and wavelet transform in Fig.~\ref{fig:exthird}(b). This higher-order harmonic-induced attenuation also agrees with the theoretical analyses of the nonlinear band structures in Fig.~\ref{fig:dispersion} and the spatial frequency analysis in Fig.~\ref{fig:thirdharmonic}. The experimental piezoelectric voltages in Figs.~\ref{fig:power}(c) and (d) are not strictly half-wave symmetric, which is why even-order harmonics emerge, as seen in Fig.~\ref{fig:exthird}(b). Together with the fundamental resonance-induced bandgap, this nonlinear metamaterial enables broadband vibration attenuation due to the simultaneous presence of multiple harmonics of nonlinear local resonators.

We further demonstrate the power performance of the electromechanical metamaterial beam with SECE interface circuits as the power supply for self-powered sensing applications. Given the load independence characteristic of the applied SECE interface circuit \cite{zhao2022graded} and the high input impedance of the energy management circuit during the energy accumulation phase in Fig.~\ref{fig:setup}(d), a zero load ($R_l=\infty$) condition is applied in the power measurements. The harvested power is therefore computed as the charging power from the time derivative of the storage voltage $V_r$ in steady state through slow sweep excitations. The harvested power and the storage voltage are shown in Figs.~\ref{fig:power}(a) and (b). With an increase in the base excitation, the maximum value and bandwidth of the harvested power also increase because of the nonlinear coupling effect from the SECE interface circuits. In addition to the resonant frequency range, the system can harvest power under modal frequencies. Notably, due to the dielectric loss of the storage capacitor $C_r$, there is negative harvested power when $C_r$ slowly dissipates the stored energy.  At the modal frequency of 37 Hz, the experimental and theoretical waveforms of the SECE interface circuits of 6 pairs of nonlinear local resonators are shown in Figs.~\ref{fig:power}(c) and (d). Theoretical waveforms are computed by Eq.~\eqref{eq:f_nl3} with the IFFT of the displacements and velocities of the local resonators of harmonic order H = 9. The amplitude of the theoretical piezoelectric voltage $v_p$ is well matched to the experimental values. The phase difference is due to the slightly different experimental and theoretical modal frequencies shown in Fig.~\ref{fig:bandgap}(a). With an increase in the base excitation to $\ddot{w}_b=0.5$ g, the saturation of the piezoelectric voltage due to the transistor $T_3$ can be observed from the voltage waveforms of the first and fifth pairs of nonlinear local resonators. We assume the first harmonic of the displacement and velocity still dominates the dynamics of the local resonators. Therefore, the small switching behavior observed as a result of higher harmonics in the experimental waveform is not accounted for in the theoretical modeling, which will be addressed in future studies. 

Finally, we evaluate the power consumption of the on-chip temperature and external acceleration self-powered sensing functions. Temperature detection is performed by the on-chip temperature sensor and the internal analog-to-digital converter of IN100 \cite{IN100}. At the same time, an external low-power consumption 3-axis acceleration sensor, BMA400, is attached to the clamped side of the host beam and connected to the compact IoT board through the inter-integrated circuit (I$^2$C) protocol, as shown in Fig.~\ref{fig:setup}(d), to monitor the input acceleration of the host beam. The compact IoT board is connected to the first pair of nonlinear local resonators under the base excitation of $\ddot{w}_b=0.5$ g at 37 Hz. As shown in Fig.~\ref{fig:iotpower}(a), the storage voltage $V_r$ of $C_r$ features the ``accumulation to release'' strategy of the energy management circuit in Fig.~\ref{fig:setup}(d). The storage voltage $V_r$ gradually increases before it reaches the upper voltage limit $V_{\mathrm{ON}}$ of the energy management circuit, after which the energy is converted to a stable voltage output $V_o$ for the power consumption of the IN100 Bluetooth beacon and the external acceleration sensor.  

As shown in Figs.~\ref{fig:iotpower}(b) and (c), the current consumption of the self-powered sensing functions is measured by inserting the Nordic Power Profiler Kit 2 as an ammeter through the VCC input of the Bluetooth beacon. Between continuous broadcasts with 1-second intervals, the Bluetooth beacon remains in sleep mode, consuming approximately 700 nA of current. A screenshot of the sensed Bluetooth signals of the temperature and acceleration data is shown in Fig.~\ref{fig:setup}(c), which are captured with a Bluetooth sniffer (ESP32-S3) on a remote computer. After 15.5 seconds, the energy stored in the capacitor $C_r$ is released to the Bluetooth beacon for a new round of data detection and signal transmission.

During active sensing and broadcasting, the Bluetooth beacon is first awakened and prepared for sensing tasks using its internal temperature sensor and external accelerometer. The peak current consumption, around 30 mA, occurs when sensing the 3-axis accelerations. Since the IN100 implements a state machine rather than a microcontroller unit, the state of the accelerometer registers cannot be read with conditional commands. Therefore, a 10 ms wait I$^2$C command is implemented to complete acceleration detection. Ultimately, the temperature and acceleration data are transmitted through channels 37 to 39, with a current consumption of less than 10 mA per transmission. 

The output voltage $V_o$ to the Bluetooth beacon is stabilized around 2 V during active sensing and broadcasting; therefore, the power consumption during this process, as shown in Fig.~\ref{fig:iotpower}(c), also follows the current curve. Although the maximum transient power is much larger than the harvested power indicated in Fig.~\ref{fig:power}, the total charge consumption for each self-powered sensing process is around 46.1 $\mu$C. This low charge and energy consumption can be readily compensated by the stored charge $C_r(V_{\mathrm{ON}}-V_{\mathrm{OFF}})\approx$127 $\mu$C on the storage capacitor during the energy accumulation phase. As shown in Fig.~\ref{fig:iotpower}(b), two complete sensing and transmitting cycles are successfully achieved, consuming roughly 92 $\mu$C in total. The third attempt fails due to the insufficient remaining charge on the storage capacitor $C_r$, which leads to the next round of energy accumulation. This behavior highlights that the IoT board operates in an intermittent mode rather than a continuous one; it wakes up once the predefined voltage threshold $V_{\mathrm{ON}}$ is reached, performs sensing and transmission, and then reenters the accumulation phase when $V_{\mathrm{OFF}}$ is reached. Different levels of stored energy can be realized by adjusting the energy management threshold voltages and the storage capacitance, which opens the possibility of more complex and richer applications of the EMetaNode system for self-powered sensing applications.

The remotely received sensed data  (via a Bluetooth sniffer ESP32-S3) are shown in Fig.~\ref{fig:sensing}. The maximum Z-direction acceleration of the out-of-plane movement matches the base excitation acceleration amplitude ($\ddot{w}_b=0.5$ g) well. This value can be directly used for structural health monitoring of the input flexural waves. The X- and Y-direction accelerations correspond to low in-plane vibrations of the host beam and the gravitational acceleration, respectively. In addition to the 3-axis accelerations, temperature monitoring is demonstrated, as shown by the gray solid line with circles.
In this paper, a video clip \textcolor{blue}{(\textit{demo.mp4})} recording the self-powered sensing and transmission is attached. 


\section{Conclusions}

This paper proposes an equivalent electromechanical friction-induced metamaterial for broadband vibration attenuation and self-powered sensing. On the mechanical side, we develop a ROM-HB method to efficiently and accurately model nonlinear metamaterials with general nonlinearities, including the nonlinear electromechanical coupling effect from piezoelectric interface circuits.  Notably, this work is the first to reveal the concept of electromechanical friction induced by synchronized switching-based interface circuits. This electrically induced friction not only facilitates energy harvesting but also enhances bandgap width and enables higher-harmonic-induced attenuation, contributing to superior vibration control performance. On the electrical side, we demonstrate a fully integrated electromechanical metamaterial node capable of self-powered sensing of temperature and vibration data. Experimental results verify that the system can effectively harvest power and sustain low-power wireless transmission, demonstrating its feasibility for autonomous IoT applications. Beyond the proposed system, the ROM–HB framework provides a general and scalable modeling strategy applicable to multi-degree-of-freedom and multidimensional large-scale structural systems with local nonlinearities. At the same time, the treatment of electromechanical nonlinear coupling through a generalized friction-like term makes the approach applicable to various interface circuits and transduction mechanisms that exhibit switching behaviors.  By integrating advanced power electronics-based interface circuits with mechanical metamaterials, this work provides both a computationally efficient modeling framework and a practical design strategy for developing digitized, self-sustaining mechanical systems with embedded sensing and vibration control functionalities.

\printcredits

\section*{Acknowledgment}

The authors acknowledge support from the ETH Research Grant (ETH-02 20-1), the H2020 FET-proactive project METAVEH under the grant agreement 952039, and the Research Grants Council of Hong Kong through the Junior Research Fellow Scheme (JRFS2526-5S10).

\appendix

\section{Elemental and Unit Cell Matrices}
\label{appendixA}
\renewcommand{\theequation}{A.\arabic{equation}}
\setcounter{equation}{0}

The elemental mass, stiffness, and nodal force matrices of a one-dimensional two-node beam element are given as
\begin{equation}
\mathbf{M}_e=\frac{m l_e}{420}\left[\begin{array}{cccc}
156 & 22 l_e & 54 & -13 l_e \\
22 l_e & 4 l_e^2 & 13 l_e & -3 l_e^2 \\
54 & 13 l_e & 156 & -22 l_e \\
-13 l_e & -3 l_e^2 & -22 l_e & 4 l_e^2
\end{array}\right],
\end{equation}
\begin{equation}
\mathbf{K}_e =\frac{D_0}{l_e^3}
\left[\begin{array}{cccc}
12 & 6 l_e & -12 & 6 l_e \\
6 l_e & 4 l_e^2 & -6 l_e & 2 l_e^2 \\
-12 & -6 l_e & 12 & -6 l_e \\
6 l_e & 2 l_e^2 & -6 l_e & 4 l_e^2
\end{array}\right],
\end{equation}
\begin{equation}
\mathbf{F}_e=\frac{f(t) l_e}{2}\left[ 1 \quad \frac{l_e}{6} \quad 1 \quad  -\frac{l_e}{6} \right]^\transpose,
\end{equation}
where $l_e$ is the length of the element.

The mass, stiffness, and external force matrices of a single element unit cell are given as
\begin{equation}
\resizebox{\columnwidth}{!}{$
\mathbf{M}_{\mathrm{er}}=\left[\begin{array}{ccccc}
\frac{156 m l_e}{420}+2m_r+\frac{156 m l_e}{420} & -\frac{22 m l_e^2}{420}+\frac{22 m l_e^2}{420} & 2m_r & \frac{54 m l_e}{420} & -\frac{13 m l_e^2}{420} \\
-\frac{22 m l_e^2}{420}+\frac{22 m l_e^2}{420} & \frac{4 m l_e^3}{420}+\frac{4 m l_e^3}{420} & 0 & \frac{13 m l_e^2}{420} & -\frac{3 m l_e^3}{420} \\
2m_r & 0 & 2m_r & 0 & 0 \\
\frac{54 m l_e}{420} & \frac{13 m l_e^2}{420} & 0 & \frac{156 m l_e}{420}+2m_r+\frac{156 m l_e}{420} & -\frac{22 m l_e^2}{420}+\frac{22 m l_e^2}{420} \\
-\frac{13 m l_e^2}{420} & -\frac{3 m l_e^3}{420} & 0 & -\frac{22 m l_e^2}{420}+\frac{22 m l_e^2}{420} & \frac{4 m l_e^3}{420}+\frac{4 m l_e^3}{420}
\end{array}\right],
$}
\end{equation}
\begin{equation}
\resizebox{\columnwidth}{!}{$
\mathbf{K}_{\mathrm{er}}=\left[\begin{array}{ccccc}
12 \frac{D_0}{l_e^3}+12 \frac{D_0}{l_e^3} & -6 \frac{D_0}{l_e^2}+6 \frac{D_0}{l_e^2} & 0 & -12 \frac{D_0}{l_e^3} & 6 \frac{D_0}{l_e^2} \\
-6 \frac{D_0}{l_e^2}+6 \frac{D_0}{l_e^2} & 4 \frac{D_0}{l_e}+4 \frac{D_0}{l_e} & 0 & -6 \frac{D_0}{l_e^2} & 2 \frac{D_0}{l_e} \\
0 & 0 & 2k_r & 0 & 0 \\
-12 \frac{D_0}{l_e^3} & -6 \frac{D_0}{l_e^2} & 0 & 12 \frac{D_0}{l_e^3}+12 \frac{D_0}{l_e^3} & -6 \frac{D_0}{l_e^2}+6 \frac{D_0}{l_e^2} \\
6 \frac{D_0}{l_e^2} & 2 \frac{D_0}{l_e} & 0 & -6 \frac{D_0}{l_e^2}+6 \frac{D_0}{l_e^2} & 4 \frac{D_0}{l_e}+4 \frac{D_0}{l_e}
\end{array}\right],
$}
\end{equation}
\begin{equation}
\resizebox{\columnwidth}{!}{$
\mathbf{F}_{\mathrm{er}}=\left[ -(\rho_0 l_e+2m_r) \ddot{w}_b  \quad 0 \quad -2m_r\ddot{w}_b \quad -(\rho_0 l_e+2m_r) \ddot{w}_b  \quad  0 \right]^\transpose,
$}
\end{equation}
where the ``2'' before $m_r$ and $k_r$ represents the sum of forces from two local resonators in each pair of parasitic beams. By constructing the metamaterial beam with multiple unit cells and moving the rows and columns concerning the displacements of nonlinear local resonators to the bottom and right sides of the global assembly, we obtain the matrix representation of the metamaterial beam, as shown in Eq.~\eqref{eq:fem}.

\section{Numerical Continuation}
\label{appendixB}
\renewcommand{\theequation}{B.\arabic{equation}}
\setcounter{equation}{0}

The numerical continuation of the unknown solution $\mathbf{r}_{\mathrm{N}}$  within the frequency range $\left[\omega_s,\omega_e\right]$ is carried out by a predictor-corrector scheme along the solution trajectory through arc-length parameterization \cite{seydel2009practical,krack2019harmonic}. The extended residual can be given as
\begin{equation}
\mathbf{R}(\mathbf{Q})=\left[\begin{array}{c}
\mathbf{r}_{\mathrm{N}}(\mathbf{Q}) \\
p_c(\mathbf{Q})
\end{array}\right]=\mathbf{0}, \quad \mathbf{Q}=\left[\begin{array}{ll}\hat{\mathbf{q}}_{\mathrm{N}}^{\top} & \omega\end{array}\right]^{\top}.
\end{equation}

The arc-length constraint with the continuation step size $\Delta s$ can be formulated as
\begin{equation}
p_c(\mathbf{Q})=\left(\mathbf{Q}-\mathbf{Q}_0\right)^{\mathrm{T}}\left(\mathbf{Q}-\mathbf{Q}_0\right)-\Delta s^2=0,
\end{equation}
where $\mathbf{Q}_0$ is the previously converged solution. And for the next iteration, the predictor is conceived along the tangent direction to the solution branch as
\begin{equation}
\mathbf{Q}^{\text {pre}}=\mathbf{Q}_0+\Delta s\mathbf{Q}_{\star},
\end{equation}
where the normalized tangent vector $\mathbf{Q}_{\star}$ satisfies
\begin{equation}
\left.\frac{\partial \mathbf{R}}{\partial \mathbf{Q}}\right|_{Q=Q_0}  \mathbf{Q}_{\star}=\mathbf{0}, \quad \mathbf{Q}_{\star}^{\mathrm{T}} \mathbf{Q}_{\star}=1.
\end{equation}

The analytical derivative of $\mathbf{R}$ is given in Eq.~\eqref{eq:dR}.
Treating the predicted next solution $\mathbf{Q}^{\text {pre}}$ as the initial guess enables the solution to be solved using the Newton--Raphson method.

\section{Regularization Factor}
\label{appendixC}
\renewcommand{\theequation}{C.\arabic{equation}}
\setcounter{equation}{0}

\begin{figure}[!t]
    \centering
    \includegraphics[width=1\columnwidth,page=16]{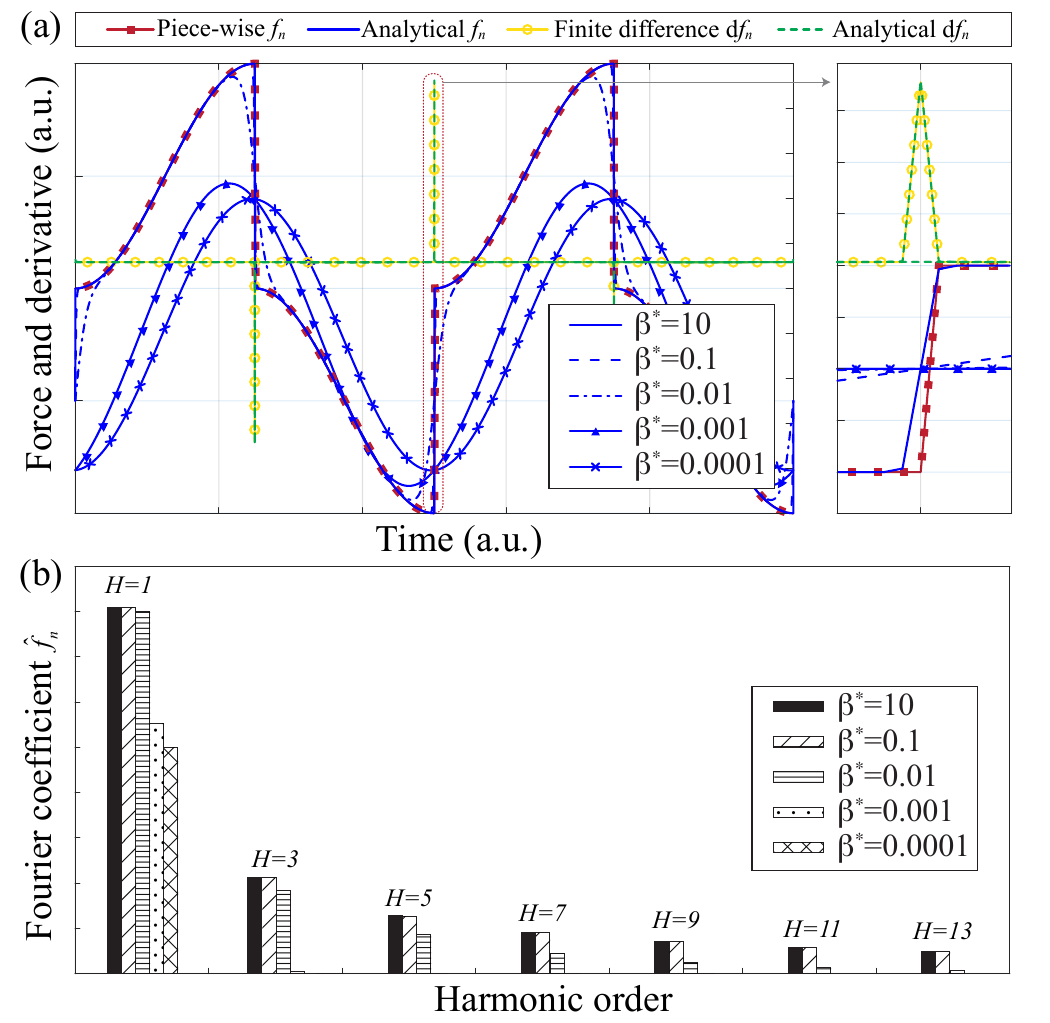}
    \caption{Influence of the regularization factor $\beta^*$. (a) Nonlinear reaction force and its derivative under different regularization factor $\beta^*$; (b) FFT of the time domain reaction force.}
    \label{fig:beta}
\end{figure}

As shown in Eq.~\eqref{eq:f_nl3}, we approximate the sign function with a hyperbolic tangent using a large regularization factor $\beta$, enabling analytical derivatives of the nonlinear reaction force due to the piezoelectric interface circuits. To assess the goodness of the regularization, we define the nondimensional factor $\beta^*=\beta\dot{U}_r$. Varying the nondimensional regularization factor results in two cases:
\begin{itemize}
    \item When $\beta^*\ll1$, $\tanh(\beta \dot{u})\approx\beta \dot{u}$, the nonlinear friction force reduces to a weak viscous damping force that is negligible compared with the elastic restoring force. Therefore, the total reaction force, as shown in Fig. \ref{fig:beta}(a), is nearly sinusoidal, and the spectrum in Fig. \ref{fig:beta}(b) is dominated by the fundamental harmonic, with vanishing higher-order harmonics. In this case, the bandgap essentially recovers the open circuit condition, as shown in Fig. \ref{fig:bandgap}(a). 
    \item When $\beta^*\gg1$, the nonlinearity saturates and $\tanh(\beta \dot{u})\approx\mathrm{sign}(\dot{u})$, yielding a Coulomb-like friction force. The resulting total nonlinear force injects odd harmonics in Fig. \ref{fig:beta}(b), which is the regime responsible for the bandgap broadening observed in Fig. \ref{fig:bandgap}(b).
\end{itemize}

The error bound of the approximation of the friction effect can be built as follows
\begin{equation}
\begin{aligned}
   \mathrm{sign}(\dot{u}_r)-\tanh (\beta \dot{u})&=\frac{2}{{\mathrm{e}}^{2 \beta \dot{u}_r}+1}, \quad \dot{u}_r>0;\\
\tanh (\beta \dot{u}_r)-\mathrm{sign}(\dot{u}_r)&=\frac{2}{\mathrm{e}^{2 \beta\dot{u}_r}+1}, \quad  \dot{u}_r<0.
\end{aligned}
\end{equation}
Therefore, 
\begin{equation}
|\tanh (\beta \dot{u})-\mathrm{sign}(\dot{u})|=\frac{2}{\mathrm{e}^{2 \beta^*}+1}<2\mathrm{e}^{-2\beta^*},
\end{equation}
and the error of the hyperbolic tangent approximation to a sign function is exponentially small. In the solving process, we choose a large $\beta$ on the basis of the smallest velocity amplitude of the local resonators in the linear solution, such that $\beta^*>10$, which makes the regularization effectively friction-like at the operating velocity scale.

\bibliographystyle{elsarticle-num}




\end{document}